\documentclass[12pt]{article}
\usepackage{graphicx}
\newlength{\vshift}
\newlength{\hshift}

\usepackage{amssymb}
\usepackage{amsmath}
\usepackage{amsthm}
\usepackage{amsopn}
\begin{document}
\vspace*{3cm}
\begin{center}
{\bf{\Large Lorentz violation parameters and noncommutative scale }}
\vskip 4em{ {\bf S. Aghababaei$^{\dag}$} \footnote{s.aghababaei@ph.iut.ac.ir}\: ,
\: {\bf M. Haghighat$^{\dag\dag}$}\footnote{m.haghighat@shirazu.ac.ir}}
\vskip 1em
$^\dag$ Department of Physics, Isfahan University of Technology, Isfahan
84156-83111, Iran
$^{\dag\dag}$ Department of Physics, Shiraz University, Shiraz 71454, Iran
\end{center}
\vspace*{1.9cm}
\begin{abstract}
\noindent We consider the noncommutative Standard Model that contains Lorentz symmetry violation as a subset of the Standard Model extension.  We introduce a constant electromagnetic field as a background to derive mutual relations between the free parameters of both theories.  As the Lorentz violation parameters of the Standard Model extension are extensively explored in different experiments and many stringent bounds on these parameters are available, we can find new bounds on the scale of noncommutativity of the order of a few to tens of teraelectron volts. 
\end{abstract}
\newpage
\section{Introduction}
\noindent The Standard Model of particle physics has achieved a remarkable phenomenological success through the past decades, but there are still unresolved various issues. Such issues are often discussed in the context of new physics or beyond the Standard Model theories.  Meanwhile, the Standard Model of particle physics as well as many other theories for describing beyond the Standard Model respect the Lorentz symmetry that is supported by many experimental inspections.  Although in the lower energy limit the Lorentz symmetry is an almost exact symmetry of nature, it is natural to study theories involving Lorentz symmetry breaking. In fact, in the Planck scale, the Lorentz symmetry violation arises through quantum gravity.  However, irrespective of the underlying fundamental theory, there is an appropriate prescription for considering both Lorentz and Charge conjugation-Parity-Time reversal (CPT) violation in the minimal Standard Model \cite{SME}.  In the so-called Standard Model extension (SME) the Lorentz violation is assumed to be induced by a spontaneous  Lorentz symmetry breaking.  Therefore, the Lorentz violated terms in the SME contain Lorentz violated (LV) parameters that are Lorentz quantities and act as constant backgrounds.  Furthermore, the SME preserves the observer Lorentz symmetry, whereas the particle Lorentz symmetry is violated.  Meanwhile, the phenomenological aspects of the SME have been extensively considered by many authors \cite{SME-ph} that have been led to very tight bounds on the LV parameters \cite{data}.  Furthermore, noncommutative (NC) space-time intrinsically breaks the Lorentz symmetry that in many works has resulted from considering the NC effects on the deviation in Lorentz symmetry invariance \cite{NC-LV}.
Moreover, the Lorentz symmetry violation in the noncommutative Standard Model (NCSM) may be systematically compared with the SME to find various relations between the LV parameters and the parameter of noncommutativity $\theta_{\mu\nu}$.  Although the NC field theories and their phenomenological aspects have been studied for
many years \cite{NC}, the obtained tight bounds on the LV parameters can provide new bounds on the value and even the components of $\theta_{\mu\nu}$.  Actually, such relations in the QED and Higgs parts of NCSM and the SME have resulted in more restricted bounds on the NC parameter \cite{NCLV, NCHiggs}. 
Here, we will study the Lorentz violation in the electroweak part of NCSM to find the corresponding relations between the LV and NC parameters. 
\noindent In this article, we briefly introduce the Lagrangian of the SME and NCSM, respectively, in Secs. 2 and 3. The mutual relations among the parameters of both theories are explored in Sec. 4.  We study the components of LV parameters to find new bounds on the value and also the components of NC parameter in Sec. 5. Moreover,  we examine the time and location dependence of LV parameters to give the location dependence of the NC parameter in different experiments. In Sec. 6, we give a summary and some concluding remarks.
\section{Standard Model extension}
The Standard Model extension provides a framework for considering the violation of Lorentz symmetry via a spontaneous symmetry breaking (SSB) at a fundamental level.  Regardless of the fundamental theory, it can be constructed by taking all possible Lorentz violating terms into account that preserve the gauge symmetry of the Standard Model and to be power-counting renormalizable. 
These additional terms are combinations of the ordinary SM fields and parameters with Lorentz indices acting as constant backgrounds that lead to particle Lorentz symmetry violation \cite{SME}.  To this end, the Lagrangian density for the electroweak part of the standard model in natural units $\hbar=c=\epsilon_{0}=1$ can be introduced as follows:
\begin{eqnarray}
\mathcal{L}^{SM}_{Fermion}=\frac{1}{2}i\overline{L}_A\gamma^{\mu}\overleftrightarrow {D_{\mu}}L_A+\frac{1}{2}i\overline{R}_A\gamma^{\mu}\overleftrightarrow {D_{\mu}}R_A,
\label{LeptonSM}
\end{eqnarray}
\begin{eqnarray}
\mathcal{L}^{SM}_{Gauge}=-\frac{1}{2}Tr(W_{\mu\nu}W^{\mu\nu})-\frac{1}{4}B_{\mu\nu}B^{\mu\nu}.
\label{GaugeSM}
\end{eqnarray}
\begin{eqnarray}
\mathcal{L}^{SM}_{Higgs}=(D_{\mu}\phi)^{\dag} D^{\mu}\phi+\mu^2 \phi^{\dag} \phi -\frac{\lambda}{3!} (\phi^{\dag} \phi)^2,
\label{HiggsSM}
\end{eqnarray}
\begin{eqnarray}
\mathcal{L}^{SM}_{Yukawa}=-[(G_L)_{AB}\overline L_{A}\phi R_{B}]+H.c.,
\label{YukawaSM}
\end{eqnarray}
where $D_\mu$ denotes the appropriate covariant derivative in each term,   $ ‎A\overleftrightarrow{\partial‎_{\mu}}B‎\equiv‎ A(‎\overrightarrow{\partial‎_{\mu}}B)-(\overrightarrow{\partial‎_{\mu}}A)B$, and  $W_{\mu\nu}$ and $B_{\mu\nu}$ are the field  strengths for the gauge groups $SU(2)$ and $U(1)$ with the gauge fields $W_\mu$ and $B_\mu$, respectively.  In the Higgs and Yukawa parts, $\phi$ shows the Higgs doublet representation with coupling $ \lambda $ and  $G_L$'s are the Yukawa couplings.  Meanwhile, the left- and right-handed fermions are defined as
\begin{equation}
    {L_A} =
 \left(\begin{array}{c}  \nu_A \\  \l_A
   \end{array} \right )_L,   R_A=(l_A)_R,
\end{equation}
for leptons and 
\begin{equation} 
    {L^\prime_A} =
 \left(\begin{array}{c} U_A \\ D_A
   \end{array} \right )_L,  R^\prime_A=(Q_A)_R,
\end{equation}
for quarks where $A=1,2,3$ labels the flavors for each generation, and $Q$ is up- or down-type quarks.
Now, we add all possible Lorentz violating terms to the SM-Lagrangian that preserve the gauge symmetries and are power-counting renormalizable. 
These additional terms can be categorized into  CPT-even that preserves the CPT symmetry and CPT-odd with the CPT symmetry violation. Therefore, the SME Lagrangian can be introduced as follows: The fermion sector
\begin{eqnarray}
\mathcal{L}^{CPT-even}_{Fermion}=i\frac{1}{2}(c_L)_{\mu\nu AB}\overline{L}_{A_{}}\gamma^{\mu}\overleftrightarrow {D^{\nu}}L_{B}
+i\frac{1}{2}(c_R)_{\mu\nu AB}\overline{R}_{A}\gamma^{\mu}\overleftrightarrow {D^{\nu}}R_{B},
\label{CPTevenLepton}
\end{eqnarray}
and
\begin{eqnarray}
\mathcal{L}^{CPT-odd}_{Fermion}=-(a_L)_{\mu AB}\overline{L}_A \gamma^{\mu} L_B-
(a_R)_{\mu AB}\overline{R}_A \gamma^{\mu} R_B,
\label{CPToddLepton}
\end{eqnarray}
where the free parameters $ c_L $ and $ a_L $ show the LV parameters in the fermion sector for the leptons.  Meanwhile, the quark terms can easily be found by replacing $L$ and $R$, respectively, with $L^\prime$ and $R^\prime$ in an appropriate manner \cite{SME}.\\
The gauge sector
\begin{eqnarray}
\mathcal{L}^{CPT-even}_{Gauge}=-\frac{1}{2}(k_W)_{\mu\nu\rho\sigma}Tr(W^{\mu\nu}W^{\rho\sigma})
-\frac{1}{4}(k_F)_{\mu\nu\rho\sigma}B^{\mu\nu}B^{\rho\sigma},
\label{CPTevenGauge}
\end{eqnarray}
and
\begin{eqnarray}
\mathcal{L}^{CPT-odd}_{Gauge}&=&(k_2)_\kappa \epsilon^{\kappa\lambda\mu\nu} Tr(W_\lambda W_{\mu\nu}+\frac{2}{3}igW_\lambda W_\mu W_\nu)\nonumber\\
&+&(k_1)_\kappa \epsilon^{\kappa\lambda\mu\nu} B_\lambda B_{\mu\nu}+(k_0)_\kappa W^\kappa,
\label{CPToddGauge}
\end{eqnarray}
where the LV parameters in this sector are $k_W$, $k_F$, and $ k_{0,1,2} $ and $Tr$ mean the trace with respect to the SU(2) group.  Nevertheless, the real parameters $ k_{0,1,2} $ are associated with a negative contribution to the energy, which leads to some instability in the SME, and one may assume them to be zero.\\
The Higgs sector
\begin{eqnarray}
\mathcal{L}^{CPT-even}_{Higgs}&=&\frac{1}{2}(k_{\phi\phi})^{\mu\nu}( D_{\mu} \phi^{\dag}) D_{\nu}\phi+H.c.\nonumber \\
&-&\frac{1}{2}(k_{\phi B})^{\mu\nu} \phi^{\dag} \phi B_{\mu\nu}-\frac{1}{2}(k_{\phi W})^{\mu\nu} \phi^{\dag} W_{\mu\nu} \phi,
\label{CPTevenHiggs}
\end{eqnarray}
for CPT-even and for CPT-odd one has
\begin{eqnarray}
\mathcal{L}^{CPT-odd}_{Higgs}=i (k_{\phi})^{\mu}\phi^{\dag} D_{\mu}\phi +H.c.,
\label{CPToddHiggs}
\end{eqnarray}
where H.c  shows  the Hermitian conjugate and $k_{\phi\phi}$, $k_{\phi B}$, $k_{\phi W}$ and $k_{\phi}$ denote the LV parameters in the Higgs sector.  Finally, the Yukawa sector can be cast into \cite{SME,Yukawaterms}
\begin{eqnarray}
\mathcal L^{CPT-even}_{Yukawa}&=&\frac{1}{2}(K_{L})_{\mu\nu}\partial^{\mu}\phi\partial^{\nu}\phi
-(h)_{AB}\overline{L}_{A}\phi R_{B}-i\gamma_5(h')_{AB}\overline{L}_{A}\phi R_{B}\nonumber\\
&-&\frac{1}{2}(H_L)_{\mu\nu AB}\overline{L}_{A}\phi \sigma^{\mu\nu}R_{B}+ H.c.,
\label{CPTevenYukawa}
\end{eqnarray}
for the CPT-even part, and the CPT-odd part can be written as 
\begin{eqnarray}
\mathcal L^{CPT-odd}_{Yukawa}&=&-(I_L)_{\mu AB}\overline{L}_{A}\gamma_{\mu}\phi R_{B}-(J_L)_{\mu AB}\overline{L}_{A}\gamma_5\gamma_\mu \phi R_{B}+H.c.,
\label{CPToddYukawa}
\end{eqnarray}
where $K_L$, $h$, $h'$, $ H_L $, $ I_L $, and $ J_L $ are the LV parameters in the Yukawa sector.
The LV parameters that are introduced in (\ref{CPTevenLepton})-(\ref{CPToddYukawa}) are sensitive to different experiments. The current bounds on the values of different LV parameters are available in Ref. \cite{data}. 
\section{Noncommutative Standard Model}
\makeatletter
\def\fmslash{\@ifnextchar[{\fmsl@sh}{\fmsl@sh[0mu]}}
\def\fmsl@sh[#1]#2{%
  \mathchoice
    {\@fmsl@sh\displaystyle{#1}{#2}}%
    {\@fmsl@sh\textstyle{#1}{#2}}%
    {\@fmsl@sh\scriptstyle{#1}{#2}}    {\@fmsl@sh\scriptscriptstyle{#1}{#2}}}
\def\@fmsl@sh#1#2#3{\m@th\ooalign{$\hfil#1\mkern#2/\hfil$\crcr$#1#3$}}
\makeatother
In noncommutative space-time, the coordinates are operators that in the canonical version obey a noncommutative relation as follows:
\begin{eqnarray}
[x^{\mu},^{*}x^{\nu}]\equiv x^\mu \star x^\nu - x^\nu \star x^\mu = i \theta^{\mu\nu}=\frac{i\epsilon_{\mu\nu}}{\Lambda_{NC}^2},
\label{Noncommutative}
\end{eqnarray}
\noindent where $\Lambda_{NC}$ denotes the NC scale of energy, and $\theta^{\mu \nu}$ is a real constant antisymmetric matrix that can be realized as two distinct constant vectors in a four-dimensional space-time. These constant vectors obviously violate the particle Lorentz symmetry that in turn relates the NCSM to a subset of the SME. Meanwhile, based on the Weyl-Moyal $\star$ product that can be defined as
\begin{eqnarray}
(f \star g)(x)= \left.\exp\!\left(\frac{i}{2}
     \theta^{\mu \nu}\frac{\partial}{\partial x^\mu}
     \frac{\partial}{\partial y^\nu}\right) f(x) g(y)\right|_{y \to x},
\end{eqnarray}
\noindent which in the leading order leads to
\begin{eqnarray}
f \star g = f \cdot g + \frac{i}{2}\theta^{\mu\nu}(x) \partial_{\mu} f \cdot \partial_{\nu} g + \mathcal{O}(\theta^2) ,
\label{starproduc}
\end{eqnarray}
\noindent one can construct the NCSM by two different approaches. As in the NC space only for a unitary group dose one have a closed Lie algebra for generators of the group; therefore, in the first approach the SM gauge group has been achieved through a two steps spontaneous symmetry breaking from the $U(3)\times U(2)\times U(1)$ symmetry group \cite{u3u2u1}.   In the second approach, by extending the algebra \cite{NCSM}, one can consider the $SU(n)$ gauge group via Seiberg-Witten (SW) maps \cite{S.W}. However, to find the relation between NCSM and SME, we consider the second approach in which the symmetry group, the number of particles, the couplings, and the gauge fields are the same as the SM in the commutative space.  
To this end, one can define the whole gauge potential $V_{\mu}$ in the noncommutative Standard Model as
\begin{equation}
 {V_\mu}=g' {B}_\mu(x)Y+g \sum_{a=1}^{3} W_{\mu a}(x) T^a_L
  +g_S \sum_{b=1}^{8} G_{\mu b}(x) T^b_S,
\end{equation}
\noindent where Y, $T^{a}_{L}$, and $T^{b}_{S}$ are the generators of $U(1)_Y$, $SU(2)_L$, and $SU(3)_C$ with the corresponding nonphysical gauge fields $ B_\mu $, $ W_\mu $, and $ G_\mu $, respectively. Therefore, the full NCSM action can be written as follows:
\begin{eqnarray}
S_{NCSM}&&=\int d^4x \sum_{i=1}^3 \overline{\widehat \Psi}^{(i)}_L \star i
\widehat{\fmslash D} \widehat \Psi^{(i)}_L
+\int d^4x \sum_{i=1}^3 \overline{\widehat \Psi}^{(i)}_R \star i
\widehat{\fmslash  D} \widehat \Psi^{(i)}_R \nonumber\\
&& \nonumber -\int d^4x \frac{1}{2 g'}
\mbox{{\bf tr}}_{\bf 1} \widehat
F_{\mu \nu} \star  \widehat F^{\mu \nu}
-\int d^4x \frac{1}{2 g} \mbox{{\bf tr}}_{\bf 2} \widehat
F_{\mu \nu} \star  \widehat F^{\mu \nu}\\
&&\nonumber
-\int d^4x \frac{1}{2 g_S} \mbox{{\bf tr}}_{\bf 3} \widehat
F_{\mu \nu} \star  \widehat F^{\mu \nu}
+ \int d^4x \bigg( \rho_0(\widehat D_\mu \widehat \Phi)^\dagger
\star \rho_0(\widehat D^\mu \widehat \Phi)
\\ && \nonumber
- \mu^2 \rho_0(\widehat {\Phi})^\dagger \star  \rho_0(\widehat \Phi) - \lambda
\rho_0(\widehat \Phi)^\dagger \star  \rho_0(\widehat \Phi)
\star
\rho_0(\widehat \Phi)^\dagger \star  \rho_0(\widehat \Phi)   \bigg)
\\ && \nonumber
+ \int d^4x \bigg (
-\sum_{i,j=1}^3 W^{ij} \bigg
( ( \bar{ \widehat L}^{(i)}_L \star \rho_L(\widehat \Phi))
\star  \widehat e^{(j)}_R
+ \bar {\widehat e}^{(i)}_R \star (\rho_L(\widehat \Phi)^\dagger \star \widehat
L^{(j)}_L) \bigg )
\\ && \nonumber
-\sum_{i,j=1}^3 G_u^{ij} \bigg
( ( \bar{\widehat Q}^{(i)}_L \star \rho_{\bar Q}(\widehat{\bar\Phi}))\star
\widehat u^{(j)}_R
+ \bar {\widehat u}^{(i)}_R \star
(\rho_{\bar Q}(\widehat{\bar\Phi})^\dagger
\star \widehat Q^{(j)}_L) \bigg )
\\ && 
-\sum_{i,j=1}^3 G_d^{ij} \bigg
( ( \bar{ \widehat Q}^{(i)}_L \star \rho_Q(\widehat \Phi))\star
\widehat d^{(j)}_R
+ \bar{ \widehat d}^{(i)}_R \star (\rho_Q(\widehat \Phi)^\dagger
\star \widehat Q^{(j)}_L) \bigg ) \bigg),
\label{NCSM}
\end{eqnarray}
where the hat denotes the field in the NC space that can be obtained in terms of the corresponding field in the ordinary space via the SW map.  For the fermion fields, Higgs field, gauge potentials, and the field strengths on the NC space-time the corresponding relations can be found in Refs.\cite{NCSM, LCNCSM}.  The matrices $W^{ij}$, $G^{ij}_u$, and $G^{ij}_d$ show the Yukawa couplings and $tr_i$'s, and $i=1,2,3$ show traces with respect to U(1)$_Y$, SU(2)$_L$, and SU(3)$_C$, respectively, as is defined in \cite{NCSM}.  Now we would like to explore the Lorentz violation in the electroweak part (EW) of the NCSM (NCEW). For this purpose, the electroweak part of NCSM should be expanded up to the first order of $ \theta $.  To this end, one can introduce the corresponding action as follows:
\begin{eqnarray}
S_{NCEW}&=&S^{NC}_{Fermion} +S^{NC}_{Gauge}+S^{NC}_{Higgs}+S^{NC}_{Yukawa}.
\end{eqnarray}
As (\ref{NCSM}) shows, one can easily see that
\begin{eqnarray}
S^{NC}_{Fermion}= \int d^4x \left (\sum_{A} \overline{\widehat
	\Psi}^{(A)}_{L} \star
i \fmslash{\widehat D} \widehat \Psi^{(A)}_{L}
+ \sum_{A} \overline{\widehat \Psi}^{(A)}_{R} \star i \fmslash{\widehat D}
\widehat \Psi^{(A)}_{R}\right), 
\label{Fermionic}
\end{eqnarray}
where $\widehat \Psi^{(A)}_{L}$ and $\widehat \Psi^{(A)}_{R}$ are the left-handed $SU(2)$ doublets and
 the right-handed $SU(2)$ singlets for the flavor $A$, respectively. To find the fermion part of action up to the first order of $ \theta $ the star product should be expanded in terms of $ \theta $ and the NC fields should be replaced by the ordinary fields up to the first order of $ \theta $ via the SW map. For instance, for the first generation of the lepton fields $\Psi_A=L_L,e_R  $ and up to the leading order  $\widehat \Psi_A=\Psi_A+\Psi_A^{(1)} $ where for the $i$th generation  $\Psi_A^{(1)} $ can be obtained as
 \begin{eqnarray}
  L_L^{(i)1}[{B}, W]
  &=&-\frac{1}{2} g'\theta^{\mu \nu} {B}_\mu \partial_\nu  L_L^{(i)}
    -\frac{1}{2} g \theta^{\mu \nu} W_\mu \partial_\nu L_L^{(i)}\nonumber
   \\
  &
  +&\frac{i}{4} \theta^{\mu \nu}
  \left( g'{ B}_\mu +g W_\mu\right)
\left( g'{ B}_\nu +g W_\nu\right)
    L_L^{(i)}, 
\label{L_Li}
\end{eqnarray}
for the left-handed leptons and 
\begin{eqnarray}
  e^{(i)1}_R[{ B}]&=&-\frac{1}{2}g' \theta^{\mu \nu}
  { B}_\mu \partial_\nu  e^{(i)}_R,
\label{e_Ri}
\end{eqnarray}
for the right-handed ones. Therefore, the leptonic part of the action can be rewritten as 
\begin{eqnarray} 
S^{NC}_{Lepton}&=&
\int d^4x \bigg ( \sum_{i} \left(\bar
L^{(i)}_{L}+  \bar L^{(i)1}_{L} \right)
\star
i
\left(\fmslash D^{SM} +  \fmslash \Gamma \right)
\star
\left(L^{(i)}_{L}+ L^{(i)1}_{L} \right) \nonumber\\ &+& 
\sum_{i} \left(\bar e^{(i)}_{R}+  \bar e^{(i)1}_{R} \right)
\star i
\left(\fmslash D^{SM} +  \fmslash \Gamma \right)
\star
\left( e^{(i)}_{R}+ e^{(i)1}_{R} \right)
 \bigg )  + {\cal O}(\theta^2),
\label{NCf}
\end{eqnarray}
in which for the vector potential in the NC space we have defined 
$ \widehat V_\mu=V_\mu+ i\Gamma_\mu $ where up to the first order of $\theta$ through the SW map one has
\begin{eqnarray}
\Gamma_\mu & = & i\frac{1}{4}\theta^{\alpha \beta}
     \{ g' { B}_\alpha + g W_\alpha,
     g' \partial_\beta { B}_\mu + g \partial_\beta W_\mu + g' B_{\beta \mu} +g W_{\beta \mu} \},
\label{gamma_mu}
\end{eqnarray}
with the field strengths  $ B_{\mu\nu} $ and $ W_{\mu\nu} $ corresponding to the gauge groups $ U(1) $ and $ SU(2) $, respectively.  By replacing (\ref{L_Li}), (\ref{e_Ri}), and (\ref{gamma_mu}) in (\ref{NCf}) and expanding the star product up to the first order of $\theta$ and after a little algebra one can find the lowest NC corrections on the leptonic action as follows:
\begin{eqnarray}
S^{NC}_{Lepton}&=& \int  d^4x \sum_{i}
 \bar L^{(i)}_{L} i \fmslash{D}^{SM}  L^{(i)}_{L}
 \nonumber\\  &-&\frac{1}{4} \theta^{\mu \nu}\int  d^4x \sum_{i}
 \bar L^{(i)}_{L} (g'B_{\mu \nu}+ gW_{\mu \nu})
 i \fmslash{D}^{SM} {L^{(i)}_{L}}
   \nonumber\\
 &-&\frac{1}{2}\theta^{\mu \nu}\int  d^4x \sum_{i}
 \bar L^{(i)}_{L} \gamma^\alpha
 (g'B_{\alpha \mu}+gW_{\alpha \mu}) i D^{SM}_\nu
 L^{(i)}_{L}
 \nonumber\\
 &+& \int  d^4x \sum_{i}
 \bar e^{(i)}_{R} i \fmslash{D}^{SM}  e^{(i)}_{R} \nonumber\\
 &-&\frac{1}{4} \theta^{\mu \nu}\int  d^4x \sum_{i}
 \bar e^{(i)}_{R}  g'B_{\mu \nu}
 i \fmslash{D}^{SM} e^{(i)}_{R} \nonumber\\
 &-&\frac{1}{2}\theta^{\mu \nu}\int  d^4x \sum_{i}
 \bar e^{(i)}_{R} \gamma^\alpha
 g'B_{\alpha \mu} i D^{SM}_\nu e^{(i)}_{R}  + {\cal O}(\theta^2),
  \label{NCf2}
\end{eqnarray}
where $D^{SM}$ shows the covariant derivative in the ordinary Standard Model. 
The quark part of the fermionic action has a similar structure to the leptonic part which can easily be obtained by inserting $\Psi_A=L'_A, R'_A $ for the left- and right-handed quarks with the appropriate SW map in the action (\ref{Fermionic}) \cite{NCSM}.\\
In (\ref{NCSM}) the gauge part of the EW action is
\begin{eqnarray}
S^{NC}_{Gauge}&=&-\int d^4x \frac{1}{2 g'} 
\mbox{{\bf tr}}_{\bf 1} \widehat
F_{\mu \nu} \star  \widehat F^{\mu \nu} 
-\int d^4x \frac{1}{2 g} \mbox{{\bf tr}}_{\bf 2} \widehat
F_{\mu \nu} \star  \widehat F^{\mu \nu},
\label{NCg1}
\end{eqnarray}
with the following expansion for the field strength:
\begin{eqnarray}
  \widehat F_{\mu \nu}&=&F_{\mu \nu}+ F^1_{\mu \nu} +{\cal O}(\theta^2),
\end{eqnarray}
\noindent where
\begin{eqnarray}
  F_{\mu \nu}&=&g'{B}_{\mu \nu}+g W_{\mu \nu},
\end{eqnarray}
\noindent and
\begin{eqnarray}
  F^1_{\mu \nu}&=& \frac{1}{2} \theta^{\alpha \beta} \{ F_{\mu \alpha},
  F_{\nu \beta} \} -\frac{1}{4} \theta^{\alpha \beta}
  \{ V_\alpha,(\partial_\beta+D_\beta) F_{\mu \nu} \}.
\end{eqnarray}
By inserting the field strengths up to the lowest order in (\ref{NCg1}) and expanding the star products one finds the EW gauge action up to the first order of $\theta$ in the NC space as 
\begin{eqnarray}
S^{NC}_{Gauge}&=&-\frac{1}{4} \, \int d^4x \, B_{\mu \nu} B^{ \mu \nu}
 -\frac{1}{2} \, {\rm Tr} \int d^4x \, W_{\mu \nu} W^{ \mu \nu}\nonumber\\
 &-&g \, \theta^{\mu
 	\nu} \, {\rm Tr} \int d^4x \, W_{\mu \rho} W_{\nu \sigma} W^{ \rho
 	\sigma}.
 \label{NCg2}
\end{eqnarray}
The NC action for the Higgs field in (\ref{NCSM}) is
\begin{eqnarray}\label{NChiggs0}
S^{NC}_{Higgs}&=&\int d^4x \bigg (
\rho_0\left( D_\mu \widehat \Phi \right)^\dagger \star\rho_0 \left( D^\mu  \widehat
\Phi \right)
\nonumber\\ & - & 
\mu^2  \rho_0(\widehat \Phi)^\dagger \star \rho_0( \widehat \Phi) -
\lambda ( \rho_0(\widehat \Phi)^\dagger \star \rho_0( \widehat \Phi)) \star 
(\rho_0( \widehat \Phi)^\dagger \star \rho_0(\widehat \Phi)) \bigg ), 
\end{eqnarray}
where up to the lowest order of $\theta$ one has
\begin{equation}
\rho_0(\hat \Phi)=\phi+\rho_0(\phi^1)+\mathcal{O}(\theta^2),
\end{equation}
with
\begin{eqnarray}
\rho_0(\phi^1)=-\frac{1}{2}\theta^{\alpha\beta}
(g'{ B}_\alpha+g W_\alpha) \partial_\beta \phi
+i\frac{1}{4} \theta^{\alpha \beta}
(g'{B}_\alpha+g W_\alpha) (g'{B}_\beta+g W_\beta) \phi.
\label{NCRhofield}
\end{eqnarray}
By retaining all terms in the action (\ref{NChiggs0}) at the leading order of the expansion in $ \theta $ one can easily find
\begin{eqnarray}
S^{NC}_{Higgs}&=& \int d^4x\Bigg( (D^{SM}_\mu\phi)^\dagger D^{SM \mu}\phi
-\mu^2 \phi^\dagger \phi
-\lambda (\phi^\dagger \phi) (\phi^\dagger \phi) \Bigg)
 \nonumber \\
&+&
\int d^4x \Bigg ( (D^{SM}_\mu\phi)^\dagger
\left( D^{SM \mu}\rho_0(\phi^1) + \frac{1}{2}
\theta^{\alpha \beta} \partial_\alpha V^{\mu} \partial_\beta \phi
+ \Gamma^\mu \phi \right)
 \nonumber\\ & +&
\left(D^{SM}_\mu \rho_0 (\phi^1) + \frac{1}{2}
\theta^{\alpha \beta} \partial_\alpha V_\mu \partial_\beta \phi
+ {\Gamma_\mu} \phi \right)^\dagger D^{SM \mu}\phi
 \nonumber\\ &
+&\frac{1}{4} \mu^2
\theta^{\mu \nu} \phi^\dagger (g' B_{\mu \nu} + g W_{\mu \nu}) \phi
-  \lambda i \theta^{\alpha \beta}
\phi^\dagger \phi (D^{SM}_\alpha \phi)^\dagger (D^{SM}_\beta \phi)
\Bigg) + {\cal O}(\theta^2).
\label{NChiggs}
\end{eqnarray}
For the NC Yukawa action, we only consider the leptonic part while the quark part can be obtained by replacing the corresponding fields with the leptonic ones. In this case, the action is
\begin{eqnarray}\label{NCyuk}
S^{NC}_{Yukawa}&=&\int d^4x \bigg ( 
-\sum_{i,j=1}^3 W^{ij} \bigg
( ( \bar{ \widehat L}^{(i)}_L \star \rho_L(\widehat \Phi))\star  
\widehat e^{(j)}_R
+ \bar {\widehat e}^{(i)}_R \star (\rho_L(\widehat \Phi)^\dagger \star \widehat
L^{(j)}_L)\bigg )\bigg),\nonumber\\
\end{eqnarray}
where by keeping only terms up to the first order of $\theta$ and using the appropriate representation for $ \rho_L $  \cite{NCSM} the action (\ref{NCyuk}) leads to
\begin{eqnarray}
S^{NC}_{Yukawa}&=&S^{SM}_{Yukawa}
- \int d^4x \bigg ( \sum_{i,j=1}^3 W^{ij} \bigg (
( \bar L^{i}_L \phi)   e^{1 j}_R +
( \bar L^{i}_L \rho_L(\phi^1))   e^{j}_R \nonumber\\
&+& ( \bar L^{1 i}_L \phi)   e^{j}_R +
i\frac{1}{2}\theta^{\alpha \beta} \partial_\alpha L^i_L \partial_\beta
\phi e^{j}_R
+ \bar e^{i}_R (\phi^{ \dagger} L^{1 j}_L)\nonumber\\
&+ &\bar e^{i}_R (\rho_L(\phi^1)^{ \dagger} L^{j}_L)+
\bar e^{1 i}_R (\phi^{\dagger} L^{j}_L) +
i\frac{1}{2}\theta^{\alpha \beta}
\partial_\alpha \bar e^{i}_R \partial_\beta \phi^\dagger L^j_L
 \bigg)\bigg),
 \label{NCyukawa}
\end{eqnarray}
which $S^{SM}_{Yukawa}$ is defined in (\ref{YukawaSM}) and $ L^{(1)}_{L}$ and $e^{(1)}_{R} $ are given in (\ref{L_Li}) and (\ref{e_Ri}).\\
 The total action through the NC parameter $\theta_{\mu\nu}$ violates the particle Lorentz symmetry, which can be considered as a subset of SME.  Now we are ready to explore the mutual relations between the LV parameters and the NC parameter.
\section{Lorentz violating parameters in terms of NC parameter}
\makeatletter
\def\fmslash{\@ifnextchar[{\fmsl@sh}{\fmsl@sh[0mu]}}
\def\fmsl@sh[#1]#2{%
  \mathchoice
    {\@fmsl@sh\displaystyle{#1}{#2}}%
    {\@fmsl@sh\textstyle{#1}{#2}}%
    {\@fmsl@sh\scriptstyle{#1}{#2}}%
    {\@fmsl@sh\scriptscriptstyle{#1}{#2}}}
\def\@fmsl@sh#1#2#3{\m@th\ooalign{$\hfil#1\mkern#2/\hfil$\crcr$#1#3$}}
\makeatother
In the previous section, the electroweak part of the NCSM has been introduced up to the first order of the NC parameter.  Since the parameter of noncommutativity is a constant tensor, consequently each sector of the action violates the Lorentz symmetry.  Therefore, it is sensible to have some relation between the NCSM and the SME.  In the following subsections, we explore the mutual relations among the parameters of both theories in each sector. 
\subsection{Fermion sector}
\noindent In the fermion sector, the Lagrangian density for the CPT-even part of the SME is
\begin{eqnarray}
\mathcal{L}^{CPT-even}_{F}=i\frac{1}{2}(c_L)_{\mu\nu}\overline{L}\gamma^{\mu}\overleftrightarrow {D^{\nu}}L
+i\frac{1}{2}(c_R)_{\mu\nu}\overline{R}\gamma^{\mu}\overleftrightarrow {D^{\nu}}R,
\label{CPTevenLepton1}
\end{eqnarray}
where with respect to the Left- and Right-handed fields  $ L=(\dfrac{1-\gamma_5}{2})\psi $ and $ R=(\dfrac{1+\gamma_5}{2})\psi $, one has 

\begin{eqnarray}
\mathcal{L}^{CPT-even}_{F}=i\frac{1}{2}c_{\mu\nu}\overline{\psi}\gamma^{\mu}\overleftrightarrow {D^{\nu}}\psi
-i\frac{1}{2} d_{\mu\nu} \overline{\psi}\gamma^{\mu}\gamma^{5}\overleftrightarrow {D^{\nu}}\psi,
\label{CPTevenfermion}
\end{eqnarray}
 where 
 \begin{eqnarray}
 c_{{\mu \nu }}=\frac{1}{2}\,(c_L)_{\mu\nu}+\frac{1}{2}\,(c_R)_{\mu\nu},
 \label{c_munu}
 \end{eqnarray}
 \begin{eqnarray}
 d_{{\mu \nu }}=\frac{1}{2}\,(c_L)_{\mu\nu}-\frac{1}{2}\,(c_R)_{\mu\nu}.
 \label{d_munu}
 \end{eqnarray}
 Meanwhile, the fermion part of the NCSM Lagrangian density up to the first order of $\theta$ can be written as
\begin{eqnarray}
{\cal L}^{NC}_{F}&=& i\frac{1}{2}(c_L)_{\mu\nu}[B,W]\overline{L}\gamma^{\mu}\overleftrightarrow {D^{\nu}}L
+i\frac{1}{2}(c_R)_{\mu\nu}[B]\overline{R}\gamma^{\mu}\overleftrightarrow {D^{\nu}}R + {\cal O}(\theta^2),
\label{NCf22}
\end{eqnarray}
in which in terms of the flat space metric $\eta_{\mu\nu}  $ 
\begin{eqnarray}
(c_L)_{\mu \nu}[B,W]=-\frac{1}{2}\theta^{\alpha\beta}(g'{B}_{\alpha\beta}+gW_{\alpha\beta})\eta_{\mu\nu}-{\theta^{\alpha}}_{\nu}(g'{B}_{{\mu \alpha }}+gW_{{\mu\alpha }})
,
\label{C_L}
\end{eqnarray}
\noindent and
\begin{eqnarray}
(c_R)_{\mu \nu }[B]=-\frac{1}{2}\theta^{\alpha\beta}g'{B}_{\alpha\beta}\eta_{\mu\nu}-{\theta^{\alpha}}_{\nu}g'{B}_{\mu \alpha },
\label{C_R}
\end{eqnarray}
or in a similar way as is defined in Eq. (\ref{CPTevenfermion}),
\begin{eqnarray}
c_{\mu \nu }[B,W]=-\frac{1}{4}\theta^{\alpha\beta}(2g'{B}_{\alpha\beta}+g W_{\alpha\beta})\eta_{\mu\nu}-\frac{1}{2}{\theta^{\alpha}}_{\nu}(2g'{ B}_{\mu \alpha}+gW_{\mu\alpha}),
\label{c_munu2}
\end{eqnarray}
\begin{eqnarray}
d_{\mu \nu}[W]=-\frac{1}{4}\theta^{\alpha\beta}g W_{\alpha\beta}\eta_{\mu\nu}-\frac{1}{2}{\theta^{\alpha}}_{\nu} g W_{\mu\alpha}.
\label{d_munu2}
\end{eqnarray}
It should be noted that $c_{\mu \nu }[B,W]$ and $d_{\mu \nu}[W]$ as defined in (\ref{c_munu2}) and (\ref{d_munu2}) are not the usual LV parameters $c$ and $d$, respectively.  In fact, they depend on the dynamical fields $B$ and $W$, which are the gauge fields before spontaneous symmetry breaking.  Therefore, to find the appropriate LV parameters from (\ref{c_munu2}) and (\ref{d_munu2}) one should perform the following steps:\\
1-Replace the  gauge fields $B$ and $W$ with the physical fields $A$ and $Z$ as follows:
\begin{equation}
   W^\pm_\mu=\frac{W^1_\mu \mp i W^2_\mu}{\sqrt{2}}, \quad
   Z_\mu=\frac{-g'{B}_\mu+gW^3_\mu}{\sqrt{g^2+g'^2}}
\quad \mbox{and} \quad
   {A}_\mu=\frac{g{B}_\mu+g' W^3_\mu}{\sqrt{g^2+g'^2}},
   \label{physicalfields}
 \end{equation}
and for the field strength tensors 
\begin{equation}
B_{\mu\nu}=\cos\theta_{\omega} A_{\mu\nu}-\sin\theta_{\omega}Z_{\mu\nu},
\end{equation}
  
\begin{equation}
W^{3}_{\mu\nu}=\sin\theta_{\omega} A_{\mu\nu}+\cos\theta_{\omega}Z_{\mu\nu},
\end{equation}
  
\noindent where $ \theta_{\omega} $ is the Weinberg angle. \\ 2-Introduce a background electromagnetic field $A^{b}_{\mu\nu}$ via $ A_{\mu\nu}\rightarrow A^{b}_{\mu\nu}+A_{\mu\nu} $ in (\ref{c_munu2}) and (\ref{d_munu2}), which leads to
\begin{eqnarray}
c_{{\mu \nu }}[A^{b}]=g\sin\theta_{\omega}(-\frac{3}{4}\theta^{\alpha\beta}A^{b}_{\alpha\beta}\eta_{\mu\nu}-\frac{3}{2}{\theta^{\alpha}}_{\nu}A^{b}_{\mu\alpha}),
\label{c_munu3}
\end{eqnarray}
and
\begin{eqnarray}
d_{{\mu \nu }}[A^{b}]=g\sin\theta_{\omega}(-\frac{1}{4}\theta^{\alpha\beta}A^{b}_{\alpha\beta}\eta_{\mu\nu}-\frac{1}{2}{\theta^{\alpha}}_{\nu}A^{b}_{\mu\alpha}),
\label{d_munu3}
\end{eqnarray}
\noindent where the LV parameters are obtained in term of the NC parameter through the electromagnetic background field ${A^{b}}_{\mu\nu}$.  In fact, (\ref{c_munu3}) and (\ref{d_munu3}) show that in the presence of the NC background the LV parameters in the fermion sector arise only when there is a background electromagnetic field.  Meanwhile, as the NCEW respects the CPT symmetry, the coefficients $a_{\mu}(L, R)$ in this sector are absent. 
\subsection{Gauge sector}
\noindent There are two versions for the gauge sector of the NCSM as is given in (\ref{NCg1}). The origin of freedom in the gauge sector is due to the fact that the commutator of two gauge parameters does not form a closed Lie algebra in noncommutative space.  The only exception is the fundamental representation of the U(N) group.  Therefore, for the gauge group of the Standard Model one needs to extend the algebra, which leads to an infinite number of undefined parameters.  They can be limited to the right number of fields and parameters via Seiberg-Witten maps, which themselves cannot be uniquely determined in the model.  However, these many degrees of freedom lead to a freedom in the kinetic term of the gauge fields.  In fact, gauge invariance alone is not enough to pick one of the possible choices \cite{nmNCSM}.  In the minimal noncommutative Standard Model (mNCSM), which has minimal modification with respect to the ordinary Standard Model, there is no cubic self-interaction term for photons.  In this case, where $ \rm tr_{1}Y^{3}=0 $, there is not any LV coefficient for the gauge sector from mNCSM in comparison with the SME.  However, if one uses the freedom in the choice of trace for the gauge fields, as a nonminimal version of NCSM (nmNCSM) one has \cite{nmNCSM}
\begin{eqnarray}
S_{\mbox{\tiny Gauge}}^{\mbox{\tiny nmNCEW}}
&=& S^{\mbox{\tiny mNCEW}}_{\mbox{\tiny Gauge}}
\nonumber \\
&+&{g'}^3\kappa_1{\theta^{\rho\sigma}}\hspace{-2mm}\int \hspace{-1mm}d^4x\,
\left(\frac{1}{4}{B}_{\rho\sigma}{B}_{\mu\nu}-{B}_{\mu\rho}
{B}_{\nu\sigma}\right){B}^{\mu\nu}
\nonumber \\
&+&g'g^2\kappa_2 \, \theta^{\rho\sigma}\hspace{-2mm}\int
\hspace{-1mm} d^4x
\left[(\frac{1}{4}{B}_{\rho\sigma} W^a_{\mu\nu}-
{B}_{\mu\rho} W^a_{\nu\sigma})W^{\mu\nu,a}\!+c.p.\right]
 \nonumber \\
&+&{\cal O}(\theta^2) \, ,
\label{nmNCSM}
\end{eqnarray}
where $\kappa_1$ and $\kappa_2$ are constant parameters, c.p. denotes the cyclic permutations of field strength tensors with respect to the Lorentz indices, $ S^{\mbox{\tiny mNCEW}}_{\mbox{\tiny Gauge}} $ is given in (\ref{NCg2}), and the field strengths ${B}_{\mu\nu}(={B}_{\mu\nu}Y)$ and $W_{\mu\nu}(=W_{\mu\nu}^aT_L^a)$  are
\begin{eqnarray}
{B}_{\mu\nu}&=&\partial_{\mu}{B}_{\nu}-\partial_{\nu}{ B}_{\mu}\:,\nonumber\\
W_{\mu\nu}^a &=& \partial_{\mu}W^a_{\nu}-\partial_{\nu}W^a_{\mu}
+g\;\epsilon^{abc}W^b_{\mu} W^c_{\nu}\:.\nonumber\\
\end{eqnarray}
Now, the triple photon coupling can be extracted from (\ref{nmNCSM}) by rewriting ${B}_{\mu\nu}$ and $W_{\mu\nu}$ in terms of the physical fields as
\begin{eqnarray}
{\cal L}^{nmNCGauge}_{\gamma\gamma\gamma}&=&\frac{e}{4} \sin2{\theta_\omega}\;{\rm K}_{\gamma\gamma\gamma}{\theta^{\rho\sigma}}\left(A_{\mu\nu}A_{\rho\sigma}-4A_{\mu\rho}A_{\nu\sigma}\right)A^{\mu\nu}
\, ,
\label{nmNCGauge}
\end{eqnarray}
with
\begin{eqnarray}
{\rm K}_{\gamma\gamma\gamma}&=&\frac{1}{2}\; g g'(\kappa_1 + 3 \kappa_2)
\, .
\end{eqnarray}
Meanwhile, $ ZZ\gamma $ and $ WW\gamma $ interactions can be obtained as
\begin{eqnarray}
{\cal L}^{nmNCGauge}_{ZZ\gamma}&=&\frac{e}{4} \sin2{\theta_W}\,{\rm K}_{ZZ \gamma}\,
{\theta^{\rho\sigma}}
\left[2A^{\mu\nu}\left(2Z_{\mu\rho}Z_{\nu\sigma}-Z_{\mu\nu}Z_{\rho\sigma}\right)\right.\nonumber\\
&+&\left. 8 A_{\mu\rho}Z^{\mu\nu}Z_{\nu\sigma} - A_{\rho\sigma}Z_{\mu\nu}Z^{\mu\nu}\right]\,,
\label{z}
\end{eqnarray}
with a similar expression for ${\cal L}^{nmNCGauge}_{WW\gamma}$ by replacing $Z$ by $W$ and ${\rm K}_{ZZ \gamma}$ by ${\rm K}_{WW \gamma}$ where
\begin{eqnarray}
{\rm K}_{ZZ\gamma}&=&\frac{-1}{2gg'}\; \left[{g'}^4\kappa_1 + g^2\left(g^2-2{g'}^2\right)\kappa_2\right]\,,
\end{eqnarray}
and 
\begin{eqnarray}
{\rm K}_{WW\gamma}&=&
-\frac{g}{2 g'}\left[{g'}^2+g^2\right]\kappa_2 \,.
\end{eqnarray}
Therefore, in the presence of the electromagnetic background, $ A_{\mu\nu}\rightarrow A^{b}_{\mu\nu}+A_{\mu\nu} $ in each NC Lagrangian term, which in comparison with the CPT-even gauge part of the SME as
\begin{eqnarray}
\mathcal{L}^{CPT-even}_{Gauge}=-\frac{1}{4}(k_W)_{\mu\nu\rho\sigma}W^{\mu\nu}W^{\rho\sigma}
-\frac{1}{4}(k_F)_{\mu\nu\rho\sigma}A^{\mu\nu}A^{\rho\sigma},
\label{CPTevenGaugephysical}
\end{eqnarray}
where $ W=[Z, W^{\pm}] $, one finds all appropriate LV coefficients for this sector as follows:
\begin{eqnarray}
(k_F)^{\mu\nu\rho\sigma}[A^{b}]&=&-8\varepsilon \theta^{\rho\sigma}(A^b)^{\mu\nu}+16\varepsilon \theta^{\nu\sigma}(A^b)^{\mu\rho}+32\varepsilon \theta^{\lambda\sigma}{(A^b)^{\mu}}_{\lambda}\eta^{\nu\rho},
\label{K_F}
\end{eqnarray}
where by rewriting (\ref{K_F}) as 
\begin{eqnarray}
(k_F)^{\mu\nu\rho\sigma}[A^{b}]&=&-2\varepsilon \theta^{\rho\sigma}(A^b)^{\mu\nu}+4\varepsilon \theta^{\nu\sigma}(A^b)^{\mu\rho}
+8\varepsilon \theta^{\lambda\sigma}{(A^b)^{\mu}}_{\lambda}\eta^{\nu\rho}\nonumber\\
&+&2\varepsilon \theta^{\rho\sigma}(A^b)^{\nu\mu}-4\varepsilon \theta^{\mu\sigma}(A^b)^{\nu\rho}-8\varepsilon \theta^{\lambda\sigma}{(A^b)^{\nu}}_{\lambda}\eta^{\mu\rho}\nonumber\\
&+&2\varepsilon \theta^{\sigma\rho}(A^b)^{\mu\nu}-4\varepsilon \theta^{\nu\rho}(A^b)^{\mu\sigma}-8\varepsilon \theta^{\lambda\rho}{(A^b)^{\mu}}_{\lambda}\eta^{\nu\sigma}\nonumber\\
&-&2\varepsilon \theta^{\rho\sigma}(A^b)^{\nu\mu}+4\varepsilon \theta^{\mu\rho}(A^b)^{\nu\sigma}+8\varepsilon \theta^{\lambda\rho}{(A^b)^{\nu}}_{\lambda}\eta^{\mu\sigma},
\label{K_F2}
\end{eqnarray}
one can see that the tensor $ k_F $ has the properties of the Riemann curvature tensor and zero double trace. Meanwhile, in a similar way one has
\begin{eqnarray}
(k_Z)^{\mu\nu\rho\sigma}[A^{b}]&=&8\varepsilon' \theta^{\rho\sigma}(A^b)^{\mu\nu}-16\varepsilon' \theta^{\nu\sigma}(A^b)^{\mu\rho}-32\varepsilon' \theta^{\lambda\sigma}{(A^b)^{\mu}}_{\lambda}\eta^{\nu\rho},\nonumber\\
&=&2\varepsilon' \theta^{\rho\sigma}(A^b)^{\mu\nu}-4\varepsilon' \theta^{\nu\sigma}(A^b)^{\mu\rho}
-8\varepsilon' \theta^{\lambda\sigma}{(A^b)^{\mu}}_{\lambda}\eta^{\nu\rho}\nonumber\\
&-&2\varepsilon' \theta^{\rho\sigma}(A^b)^{\nu\mu}+4\varepsilon' \theta^{\mu\sigma}(A^b)^{\nu\rho}+8\varepsilon' \theta^{\lambda\sigma}{(A^b)^{\nu}}_{\lambda}\eta^{\mu\rho}\nonumber\\
&-&2\varepsilon' \theta^{\sigma\rho}(A^b)^{\mu\nu}+4\varepsilon' \theta^{\nu\rho}(A^b)^{\mu\sigma}+8\varepsilon' \theta^{\lambda\rho}{(A^b)^{\mu}}_{\lambda}\eta^{\nu\sigma}\nonumber\\
&+&2\varepsilon' \theta^{\rho\sigma}(A^b)^{\nu\mu}-4\varepsilon' \theta^{\mu\rho}(A^b)^{\nu\sigma}-8\varepsilon' \theta^{\lambda\rho}{(A^b)^{\nu}}_{\lambda}\eta^{\mu\sigma},
\label{K_Z}
\end{eqnarray}
and
\begin{eqnarray}
(k_{W^{\pm}})^{\mu\nu\rho\sigma}[A^{b}]&=&8\varepsilon'' \theta^{\rho\sigma}(A^b)^{\mu\nu}-16\varepsilon'' \theta^{\nu\sigma}(A^b)^{\mu\rho}-32\varepsilon'' \theta^{\lambda\sigma}{(A^b)^{\mu}}_{\lambda}\eta^{\nu\rho},\nonumber\\
&=&2\varepsilon'' \theta^{\rho\sigma}(A^b)^{\mu\nu}-4\varepsilon'' \theta^{\nu\sigma}(A^b)^{\mu\rho}
-8\varepsilon'' \theta^{\lambda\sigma}{(A^b)^{\mu}}_{\lambda}\eta^{\nu\rho}\nonumber\\
&-&2\varepsilon'' \theta^{\rho\sigma}(A^b)^{\nu\mu}+4\varepsilon'' \theta^{\mu\sigma}(A^b)^{\nu\rho}+8\varepsilon'' \theta^{\lambda\sigma}{(A^b)^{\nu}}_{\lambda}\eta^{\mu\rho}\nonumber\\
&-&2\varepsilon'' \theta^{\sigma\rho}(A^b)^{\mu\nu}+4\varepsilon'' \theta^{\nu\rho}(A^b)^{\mu\sigma}+8\varepsilon'' \theta^{\lambda\rho}{(A^b)^{\mu}}_{\lambda}\eta^{\nu\sigma}\nonumber\\
&+&2\varepsilon'' \theta^{\rho\sigma}(A^b)^{\nu\mu}-4\varepsilon'' \theta^{\mu\rho}(A^b)^{\nu\sigma}-8\varepsilon'' \theta^{\lambda\rho}{(A^b)^{\nu}}_{\lambda}\eta^{\mu\sigma}
,
\label{K_W}
\end{eqnarray}
where $\varepsilon=\frac{e}{4}\sin2\theta_{\omega}K_{\gamma\gamma\gamma}$, $\varepsilon'=\frac{e}{4}\sin2\theta_{\omega}K_{ZZ\gamma}$, and $\varepsilon''=\frac{e}{4}\sin2\theta_{\omega}K_{WW\gamma}$. One should note that by replacing $ q\rightarrow 32\varepsilon $ in (\ref{K_F}) the  result given in \cite{NCLV} for the QED part of SME can be rederived. As one expects, the $k_{AF}$ parameter in this sector is absent.  
\subsection{Higgs sector}
\noindent In this sector the NC Higgs action is
\begin{eqnarray}\label{NCHiggs}
S^{NC}_{Higgs}&=& \int d^4x\Bigg( (D_\mu\phi)^\dagger D^{ \mu}\phi
   -\mu^2 \phi^\dagger \phi
-\lambda (\phi^\dagger \phi) (\phi^\dagger \phi) \Bigg)
   \nonumber \\ &
   +&
      \int d^4x \Bigg ( (D_\mu\phi)^\dagger
   \left( D^{ \mu}\rho_o(\phi^1) + \frac{1}{2}
   \theta^{\alpha \beta} \partial_\alpha V^{\mu} \partial_\beta \phi
 + \Gamma^\mu \phi \right)
 \nonumber\\ & +&
\left(D_\mu \rho_o (\phi^1) + \frac{1}{2}
   \theta^{\alpha \beta} \partial_\alpha V_\mu \partial_\beta \phi
 + {\Gamma_\mu} \phi \right)^\dagger D^{ \mu}\phi
 \nonumber\\ &
+&\frac{1}{4} \mu^2
\theta^{\mu \nu} \phi^\dagger (g' B_{\mu \nu} + g W^L_{\mu \nu}) \phi
-  \lambda i \theta^{\alpha \beta}
\phi^\dagger \phi (D_\alpha \phi)^\dagger (D_\beta \phi)
\Bigg) + {\cal O}(\theta^2),
 \end{eqnarray}
where after a little algebra by inserting $ \Gamma_\mu $ from (\ref{gamma_mu}) and $ \rho_o(\phi^1) $ from (\ref{NCRhofield}) in (\ref{NCHiggs}) one has

\begin{eqnarray}\label{NCHiggs1}
{\cal L}^{NC}_{Higgs}&=& \bigg(-\frac{1}{2}\theta^{\alpha\nu}\partial_{\alpha}B^{\mu}-i\lambda\theta^{\mu\nu}\phi^{\dagger}\phi\bigg)(D_{\mu}\phi)^{\dagger}D_{\nu}\phi\nonumber\\
&+&\bigg(\frac{1}{4}\mu^2 \sqrt{{g'}^2+g^2} \theta^{\mu\nu}\bigg)\phi^{\dagger}\phi B_{\mu\nu}\nonumber\\
&+&\bigg(-\frac{\sqrt{2}}{2} g \mu^2 \theta_{\mu\nu}\bigg)\phi^{\dagger}W_{\mu\nu}\phi.
\end{eqnarray}
By replacing $W$ and $B$ in (\ref{NCHiggs1}) with $A$ and $Z$ and by comparing the obtained Lagrangian with (\ref{CPTevenHiggs}), one can easily read the LV parameters in this sector as
\begin{eqnarray}
(k^S_{\phi\phi})^{\mu\nu}[A^{b}]&=&- \theta^{\alpha\nu} \partial _{\alpha}A^{b\mu},
\end{eqnarray}
\begin{eqnarray}
(k^A_{\phi\phi})^{\mu\nu}[\phi]&=&-2\lambda v^{2}\theta^{\mu\nu},
\end{eqnarray}
\begin{eqnarray}
(k_{\phi B})_{\mu\nu}=-\frac{1}{2}\mu^2 \sqrt{{g'}^2+g^2} \theta_{\mu\nu},
\end{eqnarray}
and
\begin{eqnarray}
(k_{\phi W})_{\mu\nu}=-\frac{\sqrt{2}}{2} g \mu^2 \theta_{\mu\nu},
\end{eqnarray}
where we have written $ k_{\phi\phi} $ in terms of symmetric and antisymmetric parts as $ (k_{\phi\phi})[A,\phi]=(k^S_{\phi\phi})[A]+i(k^A_{\phi\phi}[\phi]) $. We also set $ <\phi^{\dagger}\phi>\equiv v^{2} $ where $v$ is the vacuum expectation value of the Higgs field after SSB.  These parameters have already been introduced in \cite{NCHiggs}. 
\subsection{Yukawa sector}
\noindent For the Yukawa sector, we first substitute (\ref{L_Li}) and (\ref{e_Ri}) in the NC Yukawa action. Then, by using the SW map given in \cite{NCSM} and after some manipulations, the action up to the first order of $\theta$ leads to
\begin{eqnarray}\label{NCyu}
S_{Yukawa}^{NCSM}&=&S^{SM}_{Yukawa}- \int d^4x  \sum_{i,j=1}^3 W^{ij} \bigg(\nonumber\\
&&[ \frac{i}{4} \theta^{\mu\nu}i{g'}B_{\mu\nu}] (\bar L^{i}_L \phi  e^{j}_R) 
+[\frac{i}{2}\theta^{\mu\nu}]( D_{\mu} \bar L^{i}_L D_{\nu} \phi   e^{j}_R)\nonumber\\ 
&&+[\frac{1}{4}\theta^{\mu\nu}{g'}B_{\nu}]( D_{\mu} \bar L^{i}_L \phi e^{j}_R) +[ \theta^{\mu\nu}(-3{g'} B_{\nu}+2g W_{\nu})](\bar L^{i}_L D_{\mu}\phi e^{j}_R)\bigg)\nonumber\\
&&+H.c.,
\end{eqnarray}
for leptons and a similar relation for quarks.  As (\ref{NCyu}) shows, only the first term in the NC corrections can be cast into a power-counting renormalizable form and can be compared with its counterpart in the SME. By comparing (\ref{NCyu}) after SSB with (\ref{CPTevenYukawa}), one can find the coupling constant $ h $ in the NC space as follows:
\begin{eqnarray}
h[A^{b}]= \frac{1}{4} \theta^{\mu\nu}g'\cos\theta_{\omega}A^{b}_{\mu\nu},
\end{eqnarray}
where $A^b_{\mu\nu}$ is not a dynamical field and it should be considered as a constant background the same as the other LV parameters.  Therefore, for a constant magnetic field about $1 G$ and $\Lambda\sim 1 TeV$, the LV-parameter $h\sim 10^{-27}$ that is minuscule is the same as the other LV parameters. 

\section{The components of LV parameters}
In the previous section, we have found the LV parameters in the electroweak part of the SME in terms of the NC parameter.  These relations can be used to find new bounds on the value of free parameters of each theory from the existing bound on the other theory.  However, the SME free parameters are extensively examined, and there are stringent bounds on each component or some combinations of LV parameters \cite{data}.  Therefore, by studying these components new bounds on the NC parameter is expected. To this end, we define the electromagnetic background $A^{b}_{\mu\nu}$ where the Lorentz indices are $T$ and $I=X, Y, Z$ as follows:
\begin{eqnarray}
A^{b}_{TI}=(A^{b}_{TX},A^{b}_{TY},A^{b}_{TZ})=(E_{X},E_{Y},E_{Z}),\nonumber\\
A^{b}_{IJ}=(A^{b}_{YZ},A^{b}_{ZX},A^{b}_{XY})=(B_{X},B_{Y},B_{Z}),
\label{componentA}
\end{eqnarray}
and for the $\theta_{\mu\nu}$  
\begin{eqnarray}
\theta_{t}=(\theta_{TX},\theta_{TY},\theta_{TZ}),\nonumber\\
\theta_{s}=(\theta_{YZ},\theta_{ZX},\theta_{XY}),
\label{componenttheta}
\end{eqnarray}
where $\theta_{t}$ and $\theta_{s}$ are the time-space  and space-space components of the NC parameter, respectively.  Consequently, for instance, $c_{XX}$ , $c_{YY}$, and $c_{ZZ}$  from (\ref{c_munu3}) are
\begin{eqnarray}
c_{XX}&=& \alpha[-\theta_{TY}E_{Y}-\theta_{TZ}E_{Z}+\theta_{YZ}B_{X}],
\end{eqnarray}
\begin{eqnarray}
c_{YY}&=& \alpha[-\theta_{TX}E_{X}-\theta_{TZ}E_{Z}-\theta_{XZ}B_{Y}],
\end{eqnarray}
\begin{eqnarray}
c_{ZZ}&=& \alpha[-\theta_{TX}E_{X}-\theta_{TY}E_{Y}+\theta_{XY}B_{Z}],
\end{eqnarray}
\noindent 
 and a suitable combination of the LV parameters for the LV experiment as $c_{Q}=c_{XX}+c_{YY}-2c_{ZZ}$,  which lead to
\begin{eqnarray}
c_{Q}&=& \alpha[\theta_{TX}E_X+\theta_{TY}E_Y-2\theta_{TZ}E_Z\nonumber\\
&+&\theta_{YZ}B_X-\theta_{XZ}B_Y-2\theta_{XY}B_Z],
\end{eqnarray}
\noindent 
where  $\alpha=-\frac{3}{2}g\sin\theta_{\omega}$. The other important components and their relevant combinations for the fermion part are given in Table 1 and for the other sectors are found in Appendix A. As Table 1 shows, the bounds on the components of $\theta_{\mu\nu}$ can easily be obtained as shown in the third column of Table 1.
\newpage
\begin{table}[h]
	\caption{The LV components in the fermion sector.  $\alpha=-\frac{3}{2}g\sin\theta_{\omega}$ and $\beta=-\frac{1}{2}g sin\theta_{\omega}$.}\label{tab:Fermion}
	\begin{center}
		\resizebox{\textwidth}{!}{ 
		\begin{tabular}{c||c|c|c}
			Parameter &NC correspondence & Bound & System  \\
			\hline
			$c^{e}_{XX}$ & $\alpha[-\theta_{TY}E_{Y}-\theta_{TZ}E_{Z}+\theta_{YZ}B_{X}]$ & $10^{-15}$
			& Astrophysics \cite{Astrophysics}   \\
			$c^{e}_{YY}$ & $\alpha[-\theta_{TX}E_{X}-\theta_{TZ}E_{Z}-\theta_{XZ}B_{Y}]$ & $10^{-15}$& Astrophysics \cite{Astrophysics}  \\
			$c^{e}_{YZ}$ & $\alpha[\theta_{TZ}E_{Y}-\theta_{XZ}B_{Z}]$ & $10^{-15}$ & Astrophysics \cite{Astrophysics} \\
			$c^{e}_{ZY}$ & $\alpha[\theta_{TY}E_{Z}+\theta_{XY}B_{Y}]$ & $10^{-16}$ & Astrophysics \cite{Astrophysics} \\
			$c^{e}_{ZZ}$ & $\alpha[-\theta_{TX}E_{X}-\theta_{TY}E_{Y}+\theta_{XY}B_{Z}]$ & $10^{-15}$ & Astrophysics \cite{Astrophysics} \\
			$c^{e}_{TT}$ & $\alpha[-\theta_{XY}B_{Z}+\theta_{XZ}B_{Y}-\theta_{YZ}B_{X}]$ & $10^{-15}$ & Collider Physics \cite{Collider} \\
			$c^{e}_{TY}$ & $\alpha[\theta_{XY}E_{X}-\theta_{YZ}E_{Z}]$ & $10^{-15}$ & Collider Physics \cite{Collider} \\
			$c^{e}_{YT}$ & $\alpha[\theta_{TX}B_{Z}-\theta_{TZ}B_{X}]$ & $10^{-15}$ & Collider Physics \cite{Collider} \\
			$c^{e}_{TZ}$ & $\alpha[\theta_{XZ}E_{X}+\theta_{YZ}E_{Y}]$ & $10^{-13}$ & Collider Physics \cite{Collider} \\
			$c^{\nu}_{TY}$ & $\alpha[\theta_{XY}E_{X}-\theta_{YZ}E_{Z}]$  & $10^{-27}$ & IceCube \cite{IceCube} \\
			$c^{e}_{YZ}+c^{e}_{ZY}$ & $\alpha[\theta_{TZ}E_{Y}-\theta_{XZ}B_{Z}+\theta_{TY}E_{Z}+\theta_{XY}B_{Y}]$& $10^{-29}$ & Comagnetometer \cite{Comag} \\
			$c^{e}_{XZ}+c^{e}_{ZX}$ & $\alpha[\theta_{TZ}E_{X}+\theta_{YZ}B_{Z}+\theta_{TX}E_{Z}+\theta_{XY}B_{X}]$ & $10^{-29}$ & Comagnetometer \cite{Comag} \\
			$c^{e}_{XY}+c^{e}_{YX}$ & $\alpha[\theta_{TY}E_X+\theta_{YZ}B_{Y}+\theta_{TX}E_{Y}-\theta_{XZ}B_{X}]$ & $10^{-29}$ & Comagnetometer \cite{Comag} \\
			$c^{e}_{XX}-c^{e}_{YY}$ & $\alpha[-\theta_{TY}E_{Y}+\theta_{YZ}B_{X}+\theta_{TX}E_{X}+\theta_{XZ}B_{Y}]$ & $10^{-29}$ & Comagnetometer \cite{Comag} \\
			$c^{n}_{Q}$& $\alpha[\theta_{YZ}B_X-\theta_{XZ}B_Y-2\theta_{XY}B_Z]$& $10^{-25}$ & Cs/Hg clock comparison \cite{ClockLV} \\
			$c^{n}_{Q}$& $\alpha[\theta_{YZ}B_X-\theta_{XZ}B_Y-2\theta_{XY}B_Z]$& $10^{-25}$ & Be clock comparison \cite{ClockLV} \\
			$d^{e}_{XX}$ & $\beta[-\theta_{TY}E_{Y}-\theta_{TZ}E_{Z}+\theta_{YZ}B_{X}] $& $10^{-14} $ & Astrophysics \cite{Astrophysics}   \\
			$d^{e}_{YZ}$ & $\beta[\theta_{TZ}E_{Y}-\theta_{XZ}B_{Z}] $& $10^{-15}$ & Astrophysics \cite{Astrophysics} \\
			$d^{e}_{ZZ}$ & $\beta[-\theta_{TX}E_{X}-\theta_{TY}E_{Y}+\theta_{XY}B_{Z}]$ & $10^{-15}$ & Astrophysics \cite{Astrophysics} \\
			$d^{e}_{TX}$ & $\beta[-\theta_{XY}E_{Y}-\theta_{XZ}E_{Z}]$& $10^{-14}$ & Astrophysics \cite{Astrophysics} \\
			$d^{e}_{TY}$ & $\beta[\theta_{XY}E_{X}-\theta_{YZ}E_{Z}]$ & $10^{-15}$ & Astrophysics \cite{Astrophysics} \\
			$d^{e}_{TZ}$ & $\beta[\theta_{XZ}E_{X}+\theta_{YZ}E_{Y}]$ & $10^{-17}$ & Astrophysics \cite{Astrophysics} \\
			\hline
		\end{tabular}}
	\end{center}
\end{table}
\noindent The Lorentz violating parameters are defined in a nonrotating frame. Since the laboratory frame rotates with the Earth's rotation, the LV components should be time and location dependent.  Therefore, similar experiments in different places should lead to some discrepancy that is caused by the noncommutativity.  To this end, one needs some relation between the nonrotating basis $(X, Y, Z)$ and the rotating one $(x,y,z)$ where $Z$ is along the north direction parallel to the Earth's axis and $z$ is normal to surface of the Earth, as follows: 
\begin{equation}
\left(\begin{array}{c} $ x$ \\$ y $\\ $z$
\end{array}
\right)
=\left(
\begin{array}{ccc}
\cos{\chi}\cos{\Omega t} & \cos{\chi}\sin{\Omega t} & -\sin{\chi} \\
-\sin{\Omega t} & \cos{\Omega t} & 0 \\
\sin{\chi}\cos{\Omega t} & \sin{\chi}\sin{\Omega t} & \cos{\chi}\\
\end{array}
\right)
\left(
\begin{array}{c}
X \\ Y \\ Z
\end{array}
\right),
\label{coordmatrix}
\end{equation}
\\
\noindent where $\Omega \simeq{2\pi}/(23h \, 56 min)$ is the Earth's sidereal rotation frequency and $\chi$ is the angle between $Z$ and $z$ \cite{Comag, location}. By transforming the timelike and spacelike vectors of the tensors $\theta_{\mu\nu}$ and $A^b_{\mu\nu}$ as are given in (\ref{componentA}) and (\ref{componenttheta}), one has
\begin{eqnarray}
&A^{b}_{yz}=B_{x}&=\cos\chi \cos\Omega t B_{X}+\cos\chi \sin\Omega t B_{Y}-\sin\chi B_{Z},\nonumber\\
&A^{b}_{zx}=B_{y}&=-\sin\Omega t B_{X}+\cos\Omega t B_{Y},\nonumber\\
&A^{b}_{xy}=B_{z}&=\sin\chi \cos\Omega t B_{X}+\sin\chi \sin\Omega t B_{Y}+\cos\chi B_{Z},
\label{Bxyz}
\end{eqnarray}
and
\begin{eqnarray}
&A^{b}_{tx}=E_{x}&=\cos\chi \cos\Omega t E_{X}+\cos\chi \sin\Omega t E_{Y}-\sin\chi E_{Z},\nonumber\\
&A^{b}_{ty}=E_{y}&=-\sin\Omega t E_{X}+\cos\Omega t E_{Y},\nonumber\\
&A^{b}_{tz}=E_{z}&=\sin\chi \cos\Omega t E_{X}+\sin\chi \sin\Omega t E_{Y}+\cos\chi E_{Z},
\label{Exyz}
\end{eqnarray}
with similar relations for the $\theta_{\mu\nu}$ components.  Meanwhile, the components of LV parameters depend on the time and location via  $\theta_{\mu\nu}$ and $A^b_{\mu\nu}$ dependency. For instance, the combination $c_{YZ}+c_{ZY}$ leads to
\begin{eqnarray}
c_{YZ}+c_{ZY}&=&\alpha\{-\theta_{tx}E_{x}\sin2\chi \sin\Omega t -\theta_{tx}E_{y}\sin\chi\cos\Omega t - \theta_{tx}E_{z}(\cos^2\chi-\sin^2\chi)\sin\Omega t\nonumber\\
&+&\theta_{tz}E_{x}(\cos^2\chi-\sin^2\chi)\sin\Omega t+\theta_{tz}E_{y}\cos\chi\cos\Omega t+\theta_{tz}E_{z}\sin2\chi\sin\Omega t\nonumber\\
&-&\theta_{ty}E_{x}\sin\chi\cos\Omega t+\theta_{ty}E_{z}\cos\chi\cos\Omega t+\theta_{yz}B_{x}\sin2\chi\sin\Omega t\nonumber\\
&-&\theta_{yz}B_{z}(\cos^2\chi-\sin^2\chi)\sin\Omega t-\theta_{xy}B_{x}(\cos^2\chi-\sin^2\chi)\sin\Omega t\nonumber\\
&-&\theta_{xy}B_{z}\sin2\chi\sin\Omega t+\theta_{zx}B_{x}\sin\chi\cos\Omega t-\theta_{zx}B_{z}\cos\chi\cos\Omega t\nonumber\\
&+&\theta_{yz}B_{y}\sin\chi\cos\Omega t-\theta_{xy}B_{y}\cos\chi\cos\Omega t\},\nonumber\\
\end{eqnarray}
where its time average is zero as $ \overline{\sin\Omega t}=\overline{\cos\Omega t}=0 $.  All time and location dependence of the LV components and their relevant combinations are given in Appendixes B and C.  However, the nonzero parameters after the time averaging for the fermion and Higgs sectors are presented in Tables \ref{locationFermion} and \ref{locationHiggs}.  In Table \ref{locationFermion}, for simplicity, the NC location dependence is given in terms of the physical parameters in the rotating frame.  For this purpose, the magnetic vector in the rotating frame is $\overrightarrow{B}\equiv(A^{b}_{yz}, A^{b}_{zx}, A^{b}_{xy})$  which is obtained from (\ref{Bxyz}) and similarly for the vector $\overrightarrow{\theta}\equiv(\theta_{yz}, \theta_{zx}, \theta_{xy})$ in the same frame as
\begin{eqnarray}
\theta_1=\theta_{yz}&=&\cos\chi \cos\Omega t \theta_{YZ}+\cos\chi \sin\Omega t \theta_{ZX}-\sin\chi \theta_{XY},\nonumber\\
\theta_2=\theta_{zx}&=&-\sin\Omega t \theta_{YZ}+\cos\Omega t \theta_{ZX},\nonumber\\
\theta_3=\theta_{xy}&=&\sin\chi \cos\Omega t \theta_{YZ}+\sin\chi \sin\Omega t \theta_{ZX}+\cos\chi \theta_{XY},
\end{eqnarray}
which leads to $\overrightarrow{\theta}.\overrightarrow{B}=\theta_1A^{b}_{yz}+\theta_2A^{b}_{zx}+\theta_3A^{b}_{xy}$ in Table 2. 

\begin{table}[h]
	\caption{Location dependence of the nonzero LV parameters in the fermion sector. Here, $\alpha=-\frac{3}{2}g\sin\theta_{\omega}$ and the electric field has been ignored.}\label{locationFermion}
	\begin{center}
		\begin{tabular}{c||c}
			Parameter & NC location dependence \\
			\hline
			$\overline{c}_{TT}$  & $\alpha\{ -(\theta.B)_x-(\theta.B)_y-(\theta.B)_z\}$  \\
			$\overline{c}_{XX}$  & $\alpha\{\frac{1}{2}\cos^2\chi(\theta.B)_x+\frac{1}{2}(\theta.B)_y+\frac{1}{2}\sin^2\chi(\theta.B)_z+\frac{1}{4}\sin2\chi(\theta_{x}B_{z}+\theta_{z}B_{x})\}$  \\
			$\overline{c}_{YY}$ & $\alpha\{\frac{1}{2}\cos^2\chi(\theta.B)_x+\frac{1}{2}(\theta.B)_y+\frac{1}{2}\sin^2\chi(\theta.B)_z+\frac{1}{4}\sin2\chi(\theta_{x}B_{z}+\theta_{z}B_{x})\}$   \\
			$\overline{c}_{ZZ}$ & $\alpha\{\sin^2\chi(\theta.B)_x+\cos^2\chi(\theta.B)_z-\frac{1}{2}\sin2\chi(\theta_{x}B_{z}+\theta_{z}B_{x})\}$  \\
			$\overline{c}_{XY}$& $\alpha\{\frac{1}{2}\sin\chi(\theta\times B)_x-\frac{1}{2}\cos\chi(\theta\times B)_z\}$  \\
			$\overline{c}_{YX}$& $\alpha\{-\frac{1}{2}\sin\chi(\theta\times B)_x+\frac{1}{2}\cos\chi(\theta\times B)_z\} $\\
			$\overline{c}_{Q}$ & $ \alpha\{(\cos^2\chi-2\sin^2\chi)(\theta.B)_x+(\theta.B)_y+(\sin^2\chi-2\cos^2\chi)(\theta.B)_z$\\
			& $ +\frac{3}{2}\sin2\chi(\theta_{x}B_{z}+\theta_{z}B_{x})\}$\\	
		\end{tabular}
	\end{center}
\end{table}
\begin{table}[h]
	\caption{Location dependence of the nonzero LV parameters in the Higgs sector.  Here, $\alpha=-\frac{1}{2}\mu^2\sqrt{{g'}^2+g^2}$, $\beta=-\frac{\sqrt{2}}{2}g\mu^2$, and $\gamma=-2\lambda v^2$.}\label{locationHiggs}
	\begin{center}
		\begin{tabular}{c||c}
			Parameter & NC location dependence \\
			\hline
			$(k_{\phi B})_{TZ}$  & $\alpha\{-\sin\chi \theta_{tx}+\cos\chi \theta_{tz}\}$   \\
			$(k_{\phi B})_{XY}$  & $\alpha\{-\sin\chi \theta_{yz}+\cos\chi \theta_{xy}\}$   \\
			$(k_{\phi W})_{TZ}$ & $\beta\{-\sin\chi \theta_{tx}+\cos\chi \theta_{tz}\}$   \\
			$(k_{\phi W})_{XY}$ & $\beta\{-\sin\chi \theta_{yz}+\cos\chi \theta_{xy}\}$   \\
			$(k^A_{\phi \phi})_{TZ}$& $\gamma\{-\sin\chi \theta_{tx}+\cos\chi \theta_{tz}\}$   \\
			$(k^A_{\phi \phi})_{XY}$& $\gamma\{-\sin\chi \theta_{yz}+\cos\chi \theta_{xy}\} $ \\
		\end{tabular}
	\end{center}
\end{table}
Nevertheless, to find the bound on the components of $\theta_{\mu\nu}$, one should precisely examine how the electromagnetic background field affects the system under consideration. One of the experiments that leads to valuable bounds on the NC parameter is the clock comparison test.  In such a system:\\
1. The background electric field is usually of order of $10 V/cm\sim 10^{-22} GeV^2$, which is much smaller than the background magnetic field of order of $0.1-1 T\sim 10^{-17}-10^{-16}GeV^2$ \cite{Comag}. Therefore, the electric field can safely be ignored.\\
2. Although the parameter of noncommutativity  $\theta$ is fixed, the magnetic field rotates with respect to the fixed frame,  
 \begin{eqnarray}
 B_{X}&=&\cos\chi \cos\Omega t B_{x}-\sin\Omega t B_{y}+ \sin \chi \cos\Omega t B_{z},\nonumber\\
 B_{Y}&=&\cos\chi \sin\Omega t B_{x}+\cos\Omega t B_{y}+ \sin \chi \sin\Omega t B_{z},\nonumber\\
 B_{Z}&=&-\sin\chi B_{x}+\cos \chi B_{z},
 \end{eqnarray} 
 where $ \chi\simeq 118^{\circ} $ is the angle between the magnetic field and the Earth's axis of rotation in the Cs/Hg clock comparison test \cite{ClockLV}.  In this experiment the time average $ \bar{B}_X=\bar{B}_Y=0$ while $\bar{B}_Z=-0.88 B_{x}-0.46 B_{z}$.  Therefore, for $(B_x,B_y,B_z)=(0,0,B)$, one has $\bar{B}_Z=-0.46 B$, which puts a bound on $\theta_{XY}$ as $c_{Q}=\alpha (0.92B)\theta_{XY}\sim10^{-25} $ or $\mid\theta_{XY}\mid<(10TeV)^{-2}$ for $B\sim 1 T$.  
 For the other clock comparison tests available in Ref. \cite{data}, one can put new bounds on different components of the NC parameter as is given in Table \ref{tab:NCbound}.\\
 \begin{table}[h]
 \begin{center}
 	\caption{NC bounds from the clock comparison experiment for the fermion's LV parameters $ \mid \widetilde{b}^{e}_{J}\mid<10^{-27}GeV $ and $ \mid \widetilde{d}^{e}_{J}\mid<10^{-22}GeV $ where $ J=X,Y $ as is given in \cite{data}.  The corresponding nonzero parameters in the NC space lead to $\widetilde{b}_{J}= -m d_{JT}$ and $ \widetilde{d}_{J}= m(d_{TJ}+\frac{1}{2}d_{JT})  $ where $m$ is the mass of the fermion.}\label{tab:NCbound}
 	\resizebox{\textwidth}{!}{ 
 		\begin{tabular}{c||c|c|c}
 			Parameter & NC correspondence &Experimental bound & Bound on $\theta$ \\
 			\hline
 			$d^{e}_{XT}$ & $\beta[-\theta_{TY}B_{Z}-\theta_{TZ}B_{Y}]$ & $2\times 10^{-24}$&$\mid\theta_{TY}\mid<(10TeV)^{-2}$\\
 			$d^{e}_{YT}$& $\beta[\theta_{TX}B_Z-\theta_{TZ}B_X]$& $2\times 10^{-24}$&$\mid\theta_{TX}\mid<(10TeV)^{-2}$\\
 			$d^{e}_{TX}+\frac{1}{2}d^{e}_{XT}$& $\beta[-\theta_{XY}E_Y-\theta_{XZ}E_Z]+\frac{1}{2}\beta[-\theta_{TY}B_Z-\theta_{TZ}B_Y]$& $2\times 10^{-19}$&$\mid\theta_{TY}\mid<(10GeV)^{-2}$ \\
 			$d^{e}_{TX}+\frac{1}{2}d^{e}_{XT}$& $\beta[-\theta_{XY}E_X-\theta_{YZ}E_Z]+\frac{1}{2}\beta[\theta_{TX}B_Z-\theta_{TZ}B_X]$& $2\times 10^{-19}$ &$\mid\theta_{TX}\mid<(10GeV)^{-2}$\\
 			$c^{n}_{Q}$& $\alpha(\theta_{YZ}B_X-\theta_{XZ}B_Y-2\theta_{XY}B_Z)$& $10^{-25}$ &$\mid\theta_{XY}\mid<(10TeV)^{-2}$ \\
 		\end{tabular}}
 	\end{center}
 	\end{table}
\section{Conclusion}
We considered NCSM as a subset of SME to find the mutual relations between the parameters of both theories.  For this purpose, the electroweak part of the NCSM up to the first order of the NC parameter has been expanded by using the SW maps.   Although $\theta$-dependent terms violate particle Lorentz symmetry, except in the Higgs sector, they have not any counterparts in the SME.  Consequently, NCSM is considered in the presence of a constant electromagnetic field as a background.  Subsequently, a lot of relations between the LV parameters and the NC parameter in each sector of the SME have been found in Sec. 4. For the Yukawa sector, we found a power-counting renormalizable term that violates Lorentz symmetry and is proportional to the NC parameter.  This term in a background field about $1G$ and for $\Lambda\sim 1 TeV$ leads to the corresponding LV parameters of the order of $10^{-27}$, which is very small like the other LV parameters.  In Ref. \cite{data} the latest bounds from many precise measurements on the components or some combinations of LV parameters is collected, which led to new bounds on the components of NC parameters or some combinations of $\theta_{\mu\nu}$  components as is given in Tables \ref{tab:Fermion} and \ref{tab:NCbound}.  For instance, in the clock comparison test a bound of order $(10TeV)^{-2}$ can be found on the $|\theta_{XY}|$.  We also explored the time and location dependencies of the LV parameters to obtain the location dependence of different experiments on the NC parameter as is found in Tables \ref{locationFermion} and \ref{locationHiggs}.  

\section{Appendix}
\begin{appendix}
\section{The components of LV guage parameters}
In this appendix, we derive all the LV parameters in the gauge sector that are related to the $k_F$ in terms of  
the NC parameter and the electromagnetic background fields. 
\begin{eqnarray}
\widetilde{\kappa}_{tr}&=&-\frac{2}{3} \left[(k_F)^{TXTX}+(k_F)^{TYTY}+(k_F)^{TZTZ}\right]\nonumber\\
&=&-\frac{176}{3}\left[\theta^{TX}(A^{b})^{TX}+\theta^{TY}(A^{b})^{TY}+\theta^{TZ}(A^{b})^{TZ}\right]\nonumber\\
&=&-\frac{176}{3}\left[\theta^{TX}E^X+\theta^{TY}E^Y+\theta^{TZ}E^Z\right],
\end{eqnarray}
\begin{eqnarray}
k^1&=&(k_F)^{TYXZ}\nonumber\\
&=&-8\varepsilon\theta^{XZ}(A^{b})^{TY}+16\varepsilon\theta^{YZ}(A^{b})^{TX}\nonumber\\
&=&-8\varepsilon\theta^{XZ}E^Y+16\varepsilon\theta^{YZ}E^X,
\end{eqnarray}
\begin{eqnarray}
k^2&=&(k_F)^{TXYZ}\nonumber\\
&=&-8\varepsilon\theta^{YZ}(A^{b})^{TX}+16\varepsilon\theta^{XZ}(A^{b})^{TY}\nonumber\\
&=&-8\varepsilon\theta^{YZ}E^X+16\varepsilon\theta^{XZ}E^Y,
\end{eqnarray}
\begin{eqnarray}
k^3&=&(k_F)^{TYTY}-(k_F)^{XZXZ}\nonumber\\
&=&24\varepsilon\theta^{TY}(A^{b})^{TY}+24\varepsilon\theta^{ZX}(A^{b})^{XZ}-32\varepsilon\theta^{ZT}(A^{b})^{TZ}+32\varepsilon\theta^{YX}(A^{b})^{XY}\nonumber\\
&=&24\varepsilon\theta^{TY}E^Y-24\varepsilon\theta^{ZX}B^Y+32\varepsilon\theta^{TZ}E^Z-32\varepsilon\theta^{XY}B^Z,
\end{eqnarray}
\begin{eqnarray}
k^4&=&(k_F)^{TZTZ}-(k_F)^{XYXY}\nonumber\\
&=&24\varepsilon\theta^{TZ}(A^{b})^{TZ}-24\varepsilon\theta^{XY}(A^{b})^{XY}-32\varepsilon\theta^{YT}(A^{b})^{TY}+32\varepsilon\theta^{ZX}(A^{b})^{XZ}\nonumber\\
&=&24\varepsilon\theta^{TZ}E^Z-24\varepsilon\theta^{XY}B^Z+32\varepsilon\theta^{TY}E^Y-32\varepsilon\theta^{ZX}B^Y,
\end{eqnarray}
\begin{eqnarray}
k^5&=&(k_F)^{TXTY}+(k_F)^{XZYZ}\nonumber\\
&=&24\varepsilon\theta^{TY}(A^{b})^{TX}+24\varepsilon\theta^{YZ}(A^{b})^{XZ}-32\varepsilon\theta^{XY}(A^{b})^{XY}\nonumber\\
&=&24\varepsilon\theta^{TY}E^X-24\varepsilon\theta^{YZ}B^Y-32\varepsilon\theta^{XY}B^Z,
\end{eqnarray}
\begin{eqnarray}
k^6&=&(k_F)^{TXTZ}-(k_F)^{XYYZ}\nonumber\\
&=&-8\varepsilon\theta^{TZ}(A^{b})^{TX}-8\varepsilon\theta^{YZ}(A^{b})^{XY}\nonumber\\
&=&-8\varepsilon\theta^{TZ}E^X-8\varepsilon\theta^{YZ}B^Z,
\end{eqnarray}
\begin{eqnarray}
k^7&=&(k_F)^{TYTZ}+(k_F)^{XYXZ}\nonumber\\
&=&-8\varepsilon\theta^{TZ}(A^{b})^{TY}-8\varepsilon\theta^{XZ}(A^{b})^{XY}\nonumber\\
&=&-8\varepsilon\theta^{TZ}E^Y-8\varepsilon\theta^{XZ}B^Z,
\end{eqnarray}
\begin{eqnarray}
k^8&=&(k_F)^{TXXY}+(k_F)^{TZYZ}\nonumber\\
&=&-24\varepsilon\theta^{XY}(A^{b})^{TX}+24\varepsilon\theta^{YZ}(A^{b})^{TZ}\nonumber\\
&=&-24\varepsilon\theta^{XY}E^X+24\varepsilon\theta^{YZ}E^Z,
\end{eqnarray}
\begin{eqnarray}
k^9&=&(k_F)^{TXXZ}-(k_F)^{TYYZ}\nonumber\\
&=&8\varepsilon\theta^{XZ}(A^{b})^{TX}-8\varepsilon\theta^{YZ}(A^{b})^{TY}\nonumber\\
&=&8\varepsilon\theta^{XZ}E^X-8\varepsilon\theta^{YZ}E^Y,
\end{eqnarray}
\begin{eqnarray}
k^{10}&=&(k_F)^{TYXY}-(k_F)^{TZXZ}\nonumber\\
&=&-8\varepsilon\theta^{XY}(A^{b})^{TY}+8\varepsilon\theta^{XZ}(A^{b})^{TZ}\nonumber\\
&=&-8\varepsilon\theta^{XY}E^Y+8\varepsilon\theta^{XZ}E^Z,
\end{eqnarray}
where $\varepsilon $ has been introduced in the gauge subsection of Sec. 4.
\section{Time dependence on LV parameters}
  In Sec. 5, $\theta_{\mu\nu}$ and $A^{b}_{\mu\nu}$ were introduced in terms of their electric- and magnetic-like components and, subsequently, their time and location dependence.  Here, by using these relations we give the time dependence of the LV parameters in the fermion and Higgs sectors of the SME.
\begin{itemize}
\item Fermion sector
\end{itemize}
The time dependence of all components and some of their important combinations in the fermion sector are as follows:
\begin{eqnarray}
c_{TT}&=&\alpha\{-\theta_{yz}A^{b}_{yz}-\theta_{xy}A^{b}_{xy}-\theta_{zx}A^{b}_{zx}\},
\end{eqnarray}
\begin{eqnarray}
c_{XX}&=&\alpha\{-\theta_{tx}A^{b}_{tx}(\sin^2\chi+\cos^2\chi\sin^2\Omega t)-\frac{1}{2}\theta_{tx}A^{b}_{ty}\cos\chi\sin2\Omega t +\frac{1}{2}\theta_{tx}A^{b}_{tz}\sin2\chi\cos^2\Omega t\nonumber\\
&-&\frac{1}{2}\theta_{ty}A^{b}_{tx}\cos\chi \sin2\Omega t-\theta_{ty}A^{b}_{ty}\cos^2\Omega t-\frac{1}{2}\theta_{ty}A^{b}_{tz}\sin\chi \sin2\Omega t\nonumber\\
&+&\frac{1}{2}\theta_{tz}A^{b}_{tx}\sin2\chi \cos^2\Omega t-\frac{1}{2}\theta_{tz}A^{b}_{ty}\sin\chi \sin2\Omega t-\theta_{tz}A^{b}_{tz}(\cos^2\chi+\sin^2\chi \sin^2\Omega t)\nonumber\\
&+&\theta_{yz}A^{b}_{yz}\cos^2\chi \cos^2\Omega t-\frac{1}{2}\theta_{yz}A^{b}_{zx}\cos\chi \sin2\Omega t+\frac{1}{2}\theta_{yz}A^{b}_{xy}\sin2\chi \cos^2\Omega t\nonumber\\
&-&\frac{1}{2}\theta_{zx}A^{b}_{yz}\cos\chi \sin2\Omega t+\theta_{zx}A^{b}_{zx}\sin^2\Omega t-\frac{1}{2}\theta_{zx}A^{b}_{xy}\sin\chi \sin2\Omega t\nonumber\\
&+&\frac{1}{2}\theta_{xy}A^{b}_{yz}\sin2\chi\cos^2\Omega t-\frac{1}{2}\theta_{xy}A^{b}_{zx}\sin\chi\sin2\Omega t+\theta_{xy}A^{b}_{xy}\sin^2\chi \cos^2\Omega t\},
\end{eqnarray}
\begin{eqnarray}
c_{YY}&=&\alpha\{-\theta_{tx}A^{b}_{tx}(\sin^2\chi+\cos^2\chi\cos^2\Omega t)+\frac{1}{2}\theta_{tx}A^{b}_{ty}\cos\chi\sin2\Omega t+\frac{1}{2}\theta_{tx}A^{b}_{tz}\sin2\chi \sin^2\Omega t\nonumber\\
&&+\frac{1}{2}\theta_{ty}A^{b}_{tx}\cos\chi \sin2\Omega t-\theta_{ty}A^{b}_{ty}\sin^2\Omega t+\frac{1}{2}\theta_{ty}A^{b}_{tz}\sin\chi \sin2\Omega t\nonumber\\
&&+\frac{1}{2}\theta_{tz}A^{b}_{tx}\sin2\chi \sin^2\Omega t+\frac{1}{2}\theta_{tz}A^{b}_{ty}\sin\chi \sin2\Omega t-\theta_{tz}A^{b}_{tz}(\cos^2\chi +\sin^2\chi \cos^2\Omega t)\nonumber\\
&&+\theta_{yz}A^{b}_{yz}\cos^2\chi \sin^2\Omega t+\frac{1}{2}\theta_{yz}A^{b}_{zx}\cos\chi \sin2\Omega t+\frac{1}{2}\theta_{yz}A^{b}_{xy}\sin2\chi \sin^2\Omega t\nonumber\\
&&+\frac{1}{2}\theta_{zx}A^{b}_{yz}\cos\chi \sin2\Omega t+\theta_{zx}A^{b}_{zx}\cos^2\Omega t+\frac{1}{2}\theta_{zx}A^{b}_{xy}\sin\chi \sin2\Omega t\nonumber\\
&&+\frac{1}{2}\theta_{xy}A^{b}_{yz}\sin2\chi\sin^2\Omega t+\frac{1}{2}\theta_{xy}A^{b}_{zx}\sin\chi\sin2\Omega t+\theta_{xy}A^{b}_{xy}\sin^2\chi \sin^2\Omega t  \},
\end{eqnarray}
\begin{eqnarray}
c_{ZZ}&=&\alpha\{-\theta_{tx}A^{b}_{tx}\cos^2 \chi-\frac{1}{2}\theta_{tx}A^{b}_{tz}\sin2\chi-\theta_{ty}A^{b}_{ty}\nonumber\\
&&-\frac{1}{2}\theta_{tz}A^{b}_{tx}\sin2\chi-\theta_{tz}A^{b}_{tz}\sin^2\chi+\theta_{yz}A^{b}_{yz}\sin^2\chi\nonumber\\
&&-\frac{1}{2}\theta_{yz}A^{b}_{xy}\sin2\chi-\frac{1}{2}\theta_{xy}A^{b}_{yz}\sin2\chi+\theta_{xy}A^{b}_{xy}\cos^2\chi \},
\end{eqnarray}
\newpage
\begin{eqnarray}
c_{TX}&=&\alpha\{\theta_{yz}A^{b}_{tx}\sin2\chi \sin\Omega t +\theta_{yz}A^{b}_{ty}\sin\chi\cos\Omega t +\theta_{yz}A^{b}_{tz}(\sin^2\chi-\cos^2\chi)\sin\Omega t \nonumber\\
&&+\theta_{xy}A^{b}_{tx}(\sin^2\chi-\cos^2\chi)\sin\Omega t-\theta_{xy}A^{b}_{ty}\cos\chi\cos\Omega t-\theta_{xy}A^{b}_{tz}\sin2\chi\sin\Omega t\nonumber\\
&&+\theta_{zx}A^{b}_{tx}\sin\chi\cos\Omega t-\theta_{zx}A^{b}_{tz}\cos\chi\cos\Omega t\},
\end{eqnarray}
\begin{eqnarray}
c_{XT}&=&\alpha\{\theta_{tx}A^{b}_{yz}\sin2\chi \sin\Omega t +\theta_{ty}A^{b}_{yz}\sin\chi\cos\Omega t +\theta_{tz}A^{b}_{yz}(\sin^2\chi-\cos^2\chi)\sin\Omega t \nonumber\\
&&+\theta_{tx}A^{b}_{xy}(\sin^2\chi-\cos^2\chi)\sin\Omega t-\theta_{ty}A^{b}_{xy}\cos\chi\cos\Omega t-\theta_{tz}A^{b}_{xy}\sin2\chi\sin\Omega t\nonumber\\
&&+\theta_{tx}A^{b}_{zx}\sin\chi\cos\Omega t-\theta_{tz}A^{b}_{zx}\cos\chi\cos\Omega t\},
\end{eqnarray}
\begin{eqnarray}
c_{TY}&=&\alpha\{\theta_{yz}A^{b}_{ty}\sin\chi \sin\Omega t -\theta_{yz}A^{b}_{tz}\cos\Omega t +\theta_{xy}A^{b}_{tx}\cos\Omega t \nonumber\\
&-&\theta_{xy}A^{b}_{ty}\cos\chi\sin\Omega t-\theta_{zx}A^{b}_{tx}\sin\chi\sin\Omega t+\theta_{zx}A^{b}_{tz}\cos\chi\sin\Omega t\},
\end{eqnarray}
\begin{eqnarray}
c_{YT}&=&\alpha\{\theta_{ty}A^{b}_{yz}\sin\chi \sin\Omega t -\theta_{tz}A^{b}_{yz}\cos\Omega t +\theta_{tx}A^{b}_{xy}\cos\Omega t \nonumber\\
&-&\theta_{ty}A^{b}_{xy}\cos\chi\sin\Omega t-\theta_{tx}A^{b}_{zx}\sin\chi\sin\Omega t+\theta_{tz}A^{b}_{zx}\cos\chi\sin\Omega t\},
\end{eqnarray}
\begin{eqnarray}
c_{TZ}&=&\alpha\{\theta_{yz}A^{b}_{tx}\cos^2\chi \sin2\Omega t +\theta_{yz}A^{b}_{ty}\cos\chi(\cos^2\Omega t-\sin^2\Omega t) +\frac{1}{2}\theta_{yz}A^{b}_{tz}\sin2\chi\sin2\Omega t \nonumber\\
&+&\theta_{zx}A^{b}_{tx}\cos\chi(\cos^2\Omega t-\sin^2\Omega t)-\theta_{zx}A^{b}_{ty}\sin2\Omega t+\theta_{zx}A^{b}_{tz}\sin\chi(\cos^2\Omega t-\sin^2\Omega t)\nonumber\\
&+&\frac{1}{2}\theta_{xy}A^{b}_{tx}\sin2\chi\sin2\Omega t+\theta_{xy}A^{b}_{ty}\sin\chi(\cos^2\Omega t-\sin^2\Omega t)+\theta_{xy}A^{b}_{tz}\sin^2\chi\sin2\Omega t\},\nonumber\\
\end{eqnarray}
\begin{eqnarray}
c_{ZT}&=&\alpha\{\theta_{tx}A^{b}_{yz}\cos^2\chi \sin2\Omega t + \theta_{ty}A^{b}_{yz}\cos\chi(\cos^2\Omega t-\sin^2\Omega t)+\frac{1}{2}\theta_{tz}A^{b}_{yz}\sin2\chi\sin2\Omega t \nonumber\\
&&+\theta_{tx}A^{b}_{zx}\cos\chi(\cos^2\Omega t-\sin^2\Omega t)-\theta_{ty}A^{b}_{zx}\sin2\Omega t+\theta_{tz}A^{b}_{zx}\sin\chi(\cos^2\Omega t-\sin^2\Omega t)\nonumber\\
&&+\frac{1}{2}\theta_{tx}A^{b}_{xy}\sin2\chi\sin2\Omega t+\theta_{ty}A^{b}_{xy}\sin\chi(\cos^2\Omega t-\sin^2\Omega t)+\theta_{tz}A^{b}_{xy}\sin^2\chi\sin2\Omega t\},\nonumber\\
\end{eqnarray}
\begin{eqnarray}
c_{XY}&=&\alpha\{\frac{1}{2} \theta_{tx}A^{b}_{tx}\cos^2\chi \sin2\Omega t- \theta_{tx}A^{b}_{ty}\cos\chi\sin^2\Omega t+\frac{1}{4}\theta_{tx}A^{b}_{tz}\sin2\chi\sin2\Omega t \nonumber\\
&&+\theta_{ty}A^{b}_{tx}\cos\chi\cos^2\Omega t-\frac{1}{2}\theta_{ty}A^{b}_{ty}\sin2\Omega t+\theta_{ty}A^{b}_{tz}\sin\chi\cos2\Omega t\nonumber\\
&&+\frac{1}{4}\theta_{tz}A^{b}_{tx}\sin2\chi\sin2\Omega t-\theta_{tz}A^{b}_{ty}\sin\chi\sin^2\Omega t+\frac{1}{2}\theta_{tz}A^{b}_{tz}\sin^2\chi\sin2\Omega t\nonumber\\
&&-\frac{1}{2}\theta_{yz}A^{b}_{yz}\cos^2\chi\sin2\Omega t-\theta_{yz}A^{b}_{zx}\cos\chi\cos^2\Omega t-\frac{1}{4}\theta_{yz}A^{b}_{xy}\sin2\chi\sin2\Omega t \nonumber\\
&&+\theta_{zx}A^{b}_{yz}\cos\chi\sin^2\Omega t+\frac{1}{2}\theta_{zx}A^{b}_{zx}\sin2\Omega t+\theta_{zx}A^{b}_{xy}\sin\chi\sin^2\Omega t\nonumber\\
&&-\frac{1}{4}\theta_{xy}A^{b}_{yz}\sin2\chi\sin2\Omega t-\theta_{xy}A^{b}_{zx}\sin\chi\cos^2\Omega t-\frac{1}{2}\theta_{xy}A^{b}_{xy}\sin^2\chi\sin2\Omega t\},\nonumber\\
\end{eqnarray}
\begin{eqnarray}
c_{YX}&=&\alpha\{\frac{1}{2} \theta_{tx}A^{b}_{tx}\cos^2\chi \sin2\Omega t- \theta_{ty}A^{b}_{tx}\cos\chi\sin^2\Omega t+\frac{1}{4}\theta_{tz}A^{b}_{tx}\sin2\chi\sin2\Omega t \nonumber\\
&&+\theta_{tx}A^{b}_{ty}\cos\chi\cos^2\Omega t-\frac{1}{2}\theta_{ty}A^{b}_{ty}\sin2\Omega t+\theta_{tz}A^{b}_{ty}\sin\chi\cos2\Omega t\nonumber\\
&&+\frac{1}{4}\theta_{tx}A^{b}_{tz}\sin2\chi\sin2\Omega t-\theta_{ty}A^{b}_{tz}\sin\chi\sin^2\Omega t+\frac{1}{2}\theta_{tz}A^{b}_{tz}\sin^2\chi\sin2\Omega t\nonumber\\
&&-\frac{1}{2}\theta_{yz}A^{b}_{yz}\cos^2\chi\sin2\Omega t-\theta_{zx}A^{b}_{yz}\cos\chi\cos^2\Omega t-\frac{1}{4} \theta_{xy}A^{b}_{yz}\sin2\chi\sin2\Omega t\nonumber\\
&&+\theta_{yz}A^{b}_{zx}\cos\chi\sin^2\Omega t+\frac{1}{2}\theta_{zx}A^{b}_{zx}\sin2\Omega t+\theta_{xy}A^{b}_{zx}\sin\chi\sin^2\Omega t\nonumber\\
&&-\frac{1}{4}\theta_{yz}A^{b}_{xy}\sin2\chi\sin2\Omega t-\theta_{zx}A^{b}_{xy}\sin\chi\cos^2\Omega t-\frac{1}{2}\theta_{xy}A^{b}_{xy}\sin^2\chi\sin2\Omega t\},\nonumber\\
\end{eqnarray}
\begin{eqnarray}
c_{XZ}&=&\alpha\{-\frac{1}{2} \theta_{tx}A^{b}_{tx}\sin2\chi \cos\Omega t+ \theta_{tx}A^{b}_{ty}\sin\chi\sin\Omega t-\theta_{tx}A^{b}_{tz}\sin^2\chi\cos\Omega t \nonumber\\
&&+\theta_{tz}A^{b}_{tx}\cos^2\chi\cos\Omega t-\theta_{tz}A^{b}_{ty}\cos\chi\sin\Omega t+\frac{1}{2}\theta_{tz}A^{b}_{tz}\sin2\chi\cos\Omega t\nonumber\\
&&-\frac{1}{2}\theta_{yz}A^{b}_{yz}\sin2\chi\cos\Omega t+\theta_{yz}A^{b}_{xy}\cos^2\chi\cos\Omega t+\theta_{zx}A^{b}_{yz}\sin\chi\sin\Omega t\nonumber\\
&&-\theta_{zx}A^{b}_{xy}\cos\chi\sin\Omega t-\theta_{xy}A^{b}_{yz}\sin^2\chi\cos\Omega t+\frac{1}{2}\theta_{xy}A^{b}_{xy}\sin2\chi\cos\Omega t \},\nonumber\\
\end{eqnarray}
\begin{eqnarray}
c_{ZX}&=&\alpha\{-\frac{1}{2}\theta_{tx}A^{b}_{tx}\sin2\chi \cos\Omega t + \theta_{ty}A^{b}_{tx}\sin\chi\sin\Omega t-\theta_{tz}A^{b}_{tx}\sin^2\chi\cos\Omega t \nonumber\\
&&+\theta_{tx}A^{b}_{tz}\cos^2\chi\cos\Omega t-\theta_{ty}A^{b}_{tz}\cos\chi\sin\Omega t+\frac{1}{2}\theta_{tz}A^{b}_{tz}\sin2\chi\cos\Omega t\nonumber\\
&&-\frac{1}{2}\theta_{yz}A^{b}_{yz}\sin2\chi\cos\Omega t+\theta_{xy}A^{b}_{yz}\cos^2\chi\cos\Omega t+\theta_{yz}A^{b}_{zx}\sin\chi\sin\Omega t\nonumber\\
&&-\theta_{xy}A^{b}_{zx}\cos\chi\sin\Omega t-\theta_{yz}A^{b}_{xy}\sin^2\chi\cos\Omega t+\frac{1}{2} \theta_{xy}A^{b}_{xy}\sin2\chi\cos\Omega t\},\nonumber\\
\end{eqnarray}
\begin{eqnarray}
c_{YZ}&=&\alpha\{-\frac{1}{2}\theta_{tx}A^{b}_{tx}\sin2\chi \sin\Omega t -\theta_{tx}A^{b}_{ty}\sin\chi\cos\Omega t - \theta_{tx}A^{b}_{tz}\sin^2\chi\sin\Omega t\nonumber\\
&&+\theta_{tz}A^{b}_{tx}\cos^2\chi\sin\Omega t+\theta_{tz}A^{b}_{ty}\cos\chi\cos\Omega t+\frac{1}{2}\theta_{tz}A^{b}_{tz}\sin2\chi\sin\Omega t\nonumber\\
&&+\frac{1}{2}\theta_{yz}A^{b}_{yz}\sin2\chi\sin\Omega t-\theta_{yz}A^{b}_{xy}\cos^2\chi\sin\Omega t+\theta_{zx}A^{b}_{yz}\sin\chi\cos\Omega t\nonumber\\
&&-\theta_{zx}A^{b}_{xy}\cos\chi\cos\Omega t+\theta_{xy}A^{b}_{yz}\sin^2\chi\sin\Omega t-\frac{1}{2}\theta_{xy}A^{b}_{xy}\sin2\chi\sin\Omega t\},\nonumber\\
\end{eqnarray}
\begin{eqnarray}
c_{ZY}&=&\alpha\{-\frac{1}{2}\theta_{tx}A^{b}_{tx}\sin2\chi \sin\Omega t -\theta_{ty}A^{b}_{tx}\sin\chi\cos\Omega t  -\theta_{tz}A^{b}_{tx}\sin^2\chi\sin\Omega t  \nonumber\\
&&+\theta_{tx}A^{b}_{tz}\cos^2\chi\sin\Omega t+\theta_{ty}A^{b}_{tz}\cos\chi\cos\Omega t+\frac{1}{2}\theta_{tz}A^{b}_{tz}\sin2\chi\sin\Omega t\nonumber\\
&&+\frac{1}{2}\theta_{yz}A^{b}_{yz}\sin2\chi\sin\Omega t-\theta_{xy}A^{b}_{yz}\cos^2\chi\sin\Omega t+\theta_{yz}A^{b}_{zx}\sin\chi\cos\Omega t\nonumber\\
&&-\theta_{xy}A^{b}_{zx}\cos\chi\cos\Omega t+\theta_{yz}A^{b}_{xy}\sin^2\chi\sin\Omega t-\frac{1}{2}\theta_{xy}A^{b}_{xy}\sin2\chi\sin\Omega t\}.\nonumber\\
\end{eqnarray}
\begin{eqnarray}
c_{XY}+c_{YX}&=&\alpha\{\theta_{tx}A^{b}_{tx}\cos^2\chi \sin2\Omega t+\theta_{tx}A^{b}_{ty}\cos\chi(\cos^2\Omega t-\sin^2\Omega t)\nonumber\\
&&+\frac{1}{2}\theta_{tx}A^{b}_{tz}\sin2\chi\sin2\Omega t+\theta_{ty}A^{b}_{tx}\cos\chi(\cos^2\Omega t-\sin^2\Omega t)\nonumber\\
&&-\theta_{ty}A^{b}_{ty}\sin2\Omega t+\theta_{ty}A^{b}_{tz}\sin\chi(\cos2\Omega t-\sin^2\Omega t)\nonumber\\
&&+\frac{1}{2}\theta_{tz}A^{b}_{tx}\sin2\chi\sin2\Omega t+\theta_{tz}A^{b}_{ty}\sin\chi(\cos^2\Omega t-\sin^2\Omega t)\nonumber\\
&&+\theta_{tz}A^{b}_{tz}\sin^2\chi\sin2\Omega t-\theta_{yz}A^{b}_{yz}\cos^2\chi\sin2\Omega t\nonumber\\
&&-\theta_{yz}A^{b}_{zx}\cos\chi(\cos^2\Omega t-\sin^2\Omega t)-\frac{1}{2}\theta_{yz}A^{b}_{xy}\sin2\chi\sin2\Omega t \nonumber\\
&&-\theta_{zx}A^{b}_{yz}\cos\chi(\cos^2\Omega t-\sin^2\Omega t)+\theta_{zx}A^{b}_{zx}\sin2\Omega t\nonumber\\
&&-\theta_{zx}A^{b}_{xy}\sin\chi(\cos^2\Omega t-\sin^2\Omega t)-\frac{1}{2}\theta_{xy}A^{b}_{yz}\sin2\chi\sin2\Omega t\nonumber\\
&&-\theta_{xy}A^{b}_{zx}\sin\chi(\cos^2\Omega t-\sin^2\Omega t)-\theta_{xy}A^{b}_{xy}\sin^2\chi\sin2\Omega t\},\nonumber\\
\end{eqnarray}
\begin{eqnarray}
c_{XZ}+c_{ZX}&=&\alpha\{-\theta_{tx}A^{b}_{tx}\sin2\chi \cos\Omega t+ \theta_{tx}A^{b}_{ty}\sin\chi\sin\Omega t+\theta_{tx}A^{b}_{tz}(\cos^2\chi-\sin^2\chi)\cos\Omega t \nonumber\\
&&+\theta_{tz}A^{b}_{tx}(\cos^2\chi-\sin^2\chi)\cos\Omega t-\theta_{tz}A^{b}_{ty}\cos\chi\sin\Omega t+\theta_{tz}A^{b}_{tz}\sin2\chi\cos\Omega t\nonumber\\
&&+ \theta_{ty}A^{b}_{tx}\sin\chi\sin\Omega t-\theta_{ty}A^{b}_{tz}\cos\chi\sin\Omega t-\theta_{yz}A^{b}_{yz}\sin2\chi\cos\Omega t\nonumber\\
&&+\theta_{yz}A^{b}_{xy}\cos^2\chi\cos\Omega t+\theta_{zx}A^{b}_{yz}\sin\chi\sin\Omega t-\theta_{zx}A^{b}_{xy}\cos\chi\sin\Omega t\nonumber\\
&&+\theta_{xy}A^{b}_{yz}(\cos^2\chi-\sin^2\chi)\cos\Omega t+\theta_{xy}A^{b}_{xy}\sin2\chi\cos\Omega t \},
\end{eqnarray}
\begin{eqnarray}
c_{YZ}+c_{ZY}&=&\alpha\{-\theta_{tx}A^{b}_{tx}\sin2\chi \sin\Omega t -\theta_{tx}A^{b}_{ty}\sin\chi\cos\Omega t - \theta_{tx}A^{b}_{tz}(\cos^2\chi-\sin^2\chi)\sin\Omega t\nonumber\\
&&+\theta_{tz}A^{b}_{tx}(\cos^2\chi-\sin^2\chi)\sin\Omega t+\theta_{tz}A^{b}_{ty}\cos\chi\cos\Omega t+\theta_{tz}A^{b}_{tz}\sin2\chi\sin\Omega t\nonumber\\
&& -\theta_{ty}A^{b}_{tx}\sin\chi\cos\Omega t+\theta_{ty}A^{b}_{tz}\cos\chi\cos\Omega t+\theta_{yz}A^{b}_{yz}\sin2\chi\sin\Omega t\nonumber\\
&&-\theta_{yz}A^{b}_{xy}(\cos^2\chi-\sin^2\chi)\sin\Omega t-\theta_{xy}A^{b}_{yz}(\cos^2\chi-\sin^2\chi)\sin\Omega t\nonumber\\
&&-\theta_{xy}A^{b}_{xy}\sin2\chi\sin\Omega t+\theta_{zx}A^{b}_{yz}\sin\chi\cos\Omega t-\theta_{zx}A^{b}_{xy}\cos\chi\cos\Omega t\nonumber\\
&&+\theta_{yz}A^{b}_{zx}\sin\chi\cos\Omega t-\theta_{xy}A^{b}_{zx}\cos\chi\cos\Omega t\},\nonumber\\
\end{eqnarray}
\begin{eqnarray}
c_{XX}-c_{YY}&=&\alpha\{-\theta_{tx}A^{b}_{tx}\cos^2\chi(\cos^2\Omega t-\sin^2\Omega t)-\theta_{tx}A^{b}_{ty}\cos\chi\sin2\Omega t \nonumber\\ &&+\frac{1}{2}\theta_{tx}A^{b}_{tz}\sin2\chi(\cos^2\Omega t-\sin^2\Omega t)-\theta_{ty}A^{b}_{tx}\cos\chi \sin2\Omega t\nonumber\\
&&-\theta_{ty}A^{b}_{ty}(\cos^2\Omega t-\sin^2\Omega t)-\theta_{ty}A^{b}_{tz}\sin\chi \sin2\Omega t\nonumber\\
&&+\frac{1}{2}\theta_{tz}A^{b}_{tx}\sin2\chi (\cos^2\Omega t-\sin^2\Omega t)-\theta_{tz}A^{b}_{ty}\sin\chi \sin2\Omega t\nonumber\\
&&-\theta_{tz}A^{b}_{tz}\sin^2\chi(\cos^2\Omega t- \sin^2\Omega t)+\theta_{yz}A^{b}_{yz}\cos^2\chi( \cos^2\Omega t-\sin^2\Omega t)\nonumber\\
&&-\theta_{yz}A^{b}_{zx}\cos\chi \sin2\Omega t+\frac{1}{2}\theta_{yz}A^{b}_{xy}\sin2\chi (\cos^2\Omega t-\sin^2\Omega t)\nonumber\\
&&+\frac{1}{2}\theta_{zx}A^{b}_{yz}\cos\chi \sin2\Omega t-\theta_{zx}A^{b}_{zx}(\cos^2\Omega t-\sin^2\Omega t)\nonumber\\
&&-\theta_{zx}A^{b}_{xy}\sin\chi \sin2\Omega t+\frac{1}{2}\theta_{xy}A^{b}_{yz}\sin2\chi(\cos^2\Omega t-\sin^2\Omega t)\nonumber\\
&&-\theta_{xy}A^{b}_{zx}\sin\chi\sin2\Omega t+\theta_{xy}A^{b}_{xy}\sin^2\chi( \cos^2\Omega t-\sin^2\Omega t)\},\nonumber\\
\end{eqnarray}
\begin{eqnarray}
c_{Q}&=&\alpha\{\theta_{tx}A^{b}_{tx}(\cos^2\chi-2\sin^2\chi)+\frac{3}{2}\theta_{tx}A^{b}_{tz}\sin2\chi+\theta_{ty}A^{b}_{ty}\nonumber\\
&&+\frac{3}{2}\theta_{tz}A^{b}_{tx}\sin2\chi+\theta_{tz}A^{b}_{tz}(\sin^2\chi-2\cos^2\chi)+\theta_{yz}A^{b}_{yz}(\cos^2\chi-2\sin^2\chi)\nonumber\\
&&+\frac{3}{2}\theta_{yz}A^{b}_{xy}\sin2\chi+\theta_{zx}A^{b}_{zx}+\frac{3}{2}\theta_{xy}A^{b}_{yz}\sin2\chi+\theta_{xy}A^{b}_{xy}(\sin^2\chi-2\cos^2\chi)\},\nonumber\\
\end{eqnarray}
where  $\alpha=-\frac{3}{2}g\sin\theta_{\omega}$. The components of $d_{\mu\nu}$ are also the same as $c_{\mu\nu}$ components except for replacing $\alpha$ by $\beta$, which is equal to $-\frac{1}{2}g\sin\theta_{\omega}$.
\begin{itemize}
\item Higgs sector
\end{itemize}
In this sector the time dependence of the LV parameters are
\begin{eqnarray}
(k_{\phi B})_{TX}&=&\alpha\theta_{TX}=\alpha\{\cos\chi\cos\Omega t \theta_{tx}-\sin\Omega t \theta_{ty}+\sin\chi\cos\Omega t \theta_{tz}\},\nonumber\\
\end{eqnarray}
\begin{eqnarray}
(k_{\phi B})_{TY}&=&\alpha\theta_{TY}=\alpha\{\cos\chi\sin\Omega t \theta_{tx}-\cos\Omega t \theta_{ty}+\sin\chi\sin\Omega t \theta_{tz}\},\nonumber\\
\end{eqnarray}
\begin{eqnarray}
(k_{\phi B})_{TZ}&=&\alpha\theta_{TZ}=\alpha\{-\sin\chi \theta_{tx}+\cos\chi \theta_{tz}\},\nonumber\\
\end{eqnarray}
\begin{eqnarray}
(k_{\phi B})_{YZ}&=&\alpha\theta_{YZ}=\alpha\{\cos\chi\cos\Omega t \theta_{yz}-\sin\Omega t \theta_{zx}+\sin\chi\cos\Omega t \theta_{xy}\},\nonumber\\
\end{eqnarray}
\begin{eqnarray}
(k_{\phi B})_{ZX}&=&\alpha\theta_{ZX}=\alpha\{\cos\chi\sin\Omega t \theta_{yz}-\cos\Omega t \theta_{zx}+\sin\chi\sin\Omega t \theta_{xy}\},\nonumber\\
\end{eqnarray}
\begin{eqnarray}
(k_{\phi B})_{XY}&=&\alpha\theta_{XY}=\alpha\{-\sin\chi \theta_{yz}+\cos\chi \theta_{xy}\},\nonumber\\
\end{eqnarray}
where $\alpha=-\frac{1}{2}\mu^2\sqrt{{g'}^2+g^2}$. The components of $(k_{\phi W})_{\mu\nu}$ are the same as  $(k_{\phi B})_{\mu\nu}$ but replacing $\alpha$ with $\beta=-\frac{\sqrt{2}}{2}g\mu^2$.
\section {Location dependence of LV parameters}
The time averaging of the obtained LV parameters in Appendix B leads to their location dependencies as follows:
\begin{itemize}
\item Fermion sector
\end{itemize}
\begin{eqnarray}
\overline{c}_{TT}&=&{c}_{TT},\nonumber\\
\end{eqnarray}
\begin{eqnarray}
\overline{c}_{XX}&=&\alpha\{-\theta_{tx}A^{b}_{tx}(\sin^2\chi+\frac{1}{2}\cos^2\chi) +\frac{1}{4}\theta_{tx}A^{b}_{tz}\sin2\chi\nonumber\\
&-&\frac{1}{2}\theta_{ty}A^{b}_{ty}+\frac{1}{4}\theta_{tz}A^{b}_{tx}\sin2\chi \nonumber\\
&-&\theta_{tz}A^{b}_{tz}(\cos^2\chi+\frac{1}{2}\sin^2\chi )+\frac{1}{2}\theta_{yz}A^{b}_{yz}\cos^2\chi \nonumber\\
&+&\frac{1}{4}\theta_{yz}A^{b}_{xy}\sin2\chi +\frac{1}{2}\theta_{zx}A^{b}_{zx}\nonumber\\
&+&\frac{1}{4}\theta_{xy}A^{b}_{yz}\sin2\chi+\frac{1}{2}\theta_{xy}A^{b}_{xy}\sin^2\chi\},
\end{eqnarray}
\begin{eqnarray}
\overline{c}_{YY}&=&\alpha\{-\theta_{tx}A^{b}_{tx}(\sin^2\chi+\frac{1}{2}\cos^2\chi)+\frac{1}{4}\theta_{tx}A^{b}_{tz}\sin2\chi \nonumber\\
&&-\frac{1}{2}\theta_{ty}A^{b}_{ty}+\frac{1}{4}\theta_{tz}A^{b}_{tx}\sin2\chi \nonumber\\
&&-\theta_{tz}A^{b}_{tz}(\cos^2\chi +\frac{1}{2}\sin^2\chi )+\frac{1}{2}\theta_{yz}A^{b}_{yz}\cos^2\chi \nonumber\\
&&+\frac{1}{4}\theta_{yz}A^{b}_{xy}\sin2\chi +\frac{1}{2}\theta_{zx}A^{b}_{zx}\nonumber\\
&&+\frac{1}{4}\theta_{xy}A^{b}_{yz}\sin2\chi+\frac{1}{2}\theta_{xy}A^{b}_{xy}\sin^2\chi\},
\end{eqnarray}
\begin{eqnarray}
\overline{c}_{ZZ}&=&c_{ZZ},
\end{eqnarray}
\begin{eqnarray}
\overline{c}_{XY}&=&\alpha\{-\frac{1}{2}\theta_{tx}A^{b}_{ty}\cos\chi+\frac{1}{2}\theta_{ty}A^{b}_{tx}\cos\chi+\frac{1}{2}\theta_{ty}A^{b}_{tz}\sin\chi\nonumber\\
&&-\frac{1}{2}\theta_{tz}A^{b}_{ty}\sin\chi-\frac{1}{2}\theta_{yz}A^{b}_{zx}\cos\chi+\frac{1}{2}\theta_{zx}A^{b}_{yz}\cos\chi\nonumber\\
&&+\frac{1}{2}\theta_{zx}A^{b}_{xy}\sin\chi-\frac{1}{2}\theta_{xy}A^{b}_{zx}\sin\chi\},
\end{eqnarray}
\begin{eqnarray}
\overline{c}_{YX}&=&\alpha\{-\frac{1}{2}\theta_{ty}A^{b}_{tx}\cos\chi+\frac{1}{2}\theta_{tx}A^{b}_{ty}\cos\chi+\frac{1}{2}\theta_{tz}A^{b}_{ty}\sin\chi\nonumber\\
&&-\frac{1}{2}\theta_{ty}A^{b}_{tz}\sin\chi-\frac{1}{2}\theta_{zx}A^{b}_{yz}\cos\chi+\frac{1}{2}\theta_{yz}A^{b}_{zx}\cos\chi\nonumber\\
&&+\frac{1}{2}\theta_{xy}A^{b}_{zx}\sin\chi-\frac{1}{2}\theta_{zx}A^{b}_{xy}\sin\chi\},
\end{eqnarray}
\begin{eqnarray}
\overline{c}_{Q}&=c_{Q}.
\end{eqnarray}
It should be noted that the other components and their combinations have vanished with time averaging.
\begin{itemize}
\item Higgs sector
\end{itemize}
In the Higgs sector, the nonvanishing location dependencies can be obtained as follows:
\begin{eqnarray}
\overline{(k_{\phi B})}_{TZ}&=&{(k_{\phi B})}_{TZ},\nonumber\\
\end{eqnarray}
\begin{eqnarray}
\overline{(k_{\phi B})}_{XY}&=&{(k_{\phi B})}_{XY},\nonumber\\
\end{eqnarray}
in which $\alpha=-\frac{1}{2}\mu^2\sqrt{{g'}^2+g^2}$. The location dependence of the other Higgs coefficients $\overline{(k_{\phi W})}$ and $\overline{(k^A_{\phi \phi})}$ are the same as $\overline{(k_{\phi B})}$ except replacing $\alpha$ with  $\beta=-\frac{\sqrt{2}}{2}g\mu^2$ and $\gamma=-2\lambda v^2$, respectively.
\end{appendix}

\end{document}